\documentclass[11pt, english]{article}
\usepackage{ucs}

\usepackage{color}

\usepackage{multicol, here, graphicx, lscape, babel}
\usepackage[bf]{caption2}
\usepackage{amsmath}
\usepackage{amsfonts}
\usepackage{amssymb}

\newcommand{\weglassen}[1]{}
\newcommand{\wegPeter}[1]{}

\def\be{\begin{equation}}
\def\ee{\end{equation}}

\renewcommand{\d}{\text{d}}

\newcommand{\ga}{\boldsymbol{\gamma}}
\renewcommand{\l}{\boldsymbol{l}}
\renewcommand{\a}{\boldsymbol{a}}
\renewcommand{\b}{\boldsymbol{b}}

\renewcommand{\v}{\boldsymbol{v}}
\renewcommand{\u}{\boldsymbol{u}}

\newcommand{\om}{\boldsymbol{\omega}}
\renewcommand{\r}{\mathbf{r}}
\newcommand{\A}{A}

\newcommand{\E}{\mathcal{E}}
\newcommand{\F}{F}
\newcommand{\Ehl}{\mathcal{E}_{h,l}^3}
\newcommand{\Phl}{\mathcal{P}_{h,l}^2}
\newcommand{\Pp}{\mathcal{P}_{h,l}^+}
\newcommand{\Pm}{\mathcal{P}_{h,l}^-}
\newcommand{\Uhl}{\mathcal{U}_{h,l}}

\newcommand{\R}{{\mathbb R}}
\newcommand{\dee}{{\rm d}}

\begin{document}

\title{A Poincar\'e section for the general heavy rigid body}
\author{Sven Schmidt$^{1,2}$, Holger R. Dullin$^{1,2}$, Peter H. Richter$^2$\\
$^{1}$ Department of Mathematical Sciences,\\
 Loughborough University, LE11 3TU, UK \\
 {\small h.r.dullin@lboro.ac.uk} \\
$^{2}$ Institut f\"{u}r Theoretische Physik,\\
Universit\"{a}t Bremen, D-28334
Bremen, Germany\\
}
\date{March 3rd, 2008}

\maketitle

\begin{abstract}
A general recipe is developed for the study of rigid body dynamics in terms of Poincar\'e surfaces
of section. A section condition is chosen which captures every trajectory on a given energy
surface. The possible topological types of the corresponding surfaces of section are determined,
and their 1:1 projection to a conveniently defined torus is proposed for graphical rendering.

\vspace{0.5cm}
\end{abstract}

\section{Introduction}

The dynamics of a rigid body in a constant field of gravity, with one point held fixed,
is in general non-integrable. The configuration space SO(3) has three degrees of freedom, but
except for the special cases of Euler, Lagrange, and Kovalevskaya, the system possesses only
two integrals of motion: the energy~$H=h$ and the angular momentum component in the direction of
gravity, $L_z=l$. Liouville integrability would require a third integral, but this does not exists.
The parameter space is essentially 4-dimensional: when lengths, times, and energies are properly
scaled, there remain two freedoms for the principal moments of inertia $A = \text{diag}(A_1,A_2,A_3)$,
and two for the location of the body's center of mass~$\r=(r_1,r_2,r_3)$ relative to the fixed point.
Within this 4-D set of parameters, the Euler case $\r = 0$ defines a 2-D subset (two freedoms
in the moments of inertia), Lagrange's case a 1-D subset (one free ratio of the moments
of inertia, no freedom for the direction of~$\r$), and Kovalevskaya's case is a single point
(no freedom in the moments of inertia nor in the direction of~$\r$, provided the body-fixed frame of
reference is properly chosen). This implies that the vast majority of rigid body systems
exhibits some degree of chaotic motion which to this very day has largely remained unexplored.

The three degrees of freedom of configuration space SO(3)
may be reduced to effectively two in view of the S$^1$-symmetry with respect to the axis of gravity.
Thereby the configuration space reduces to the Poisson sphere S$^2$, and the corresponding 4-D phase
spaces T$^\ast_l\text{S}^2$ are labeled with the angular momentum constant~$l$.
The surfaces $\Ehl$ of constant energy $H=h$ are manifolds in T$^\ast_l\text{S}^2$, except
at values $(h,l)$ where their topological character undergoes a bifurcation. The first step in the
analysis of phase space structure is the identification of the topology of $\Ehl$, and of its
bifurcation scheme. This requires to study the energy-momentum map from phase space to the
$(h,l)$-plane, and to determine its critical values. Already this first step
is more difficult than one might think. It has been worked out for certain
subsets of parameters~\cite{Katok72, Tatarinov73} but not for the entire 4-D family
of rigid bodies. However, even though the bifurcation schemes have not been resolved for all
cases, it is known that the connected components of $\Ehl$ come only in four
types~\cite{Tatarinov74}: sphere S$^3$, direct product $\text{S}^1\times\text{S}^2$,
real projective space $\mathbb{R}\text{P}^3$, or connected sum
$(\text{S}^1\times\text{S}^2)\#(\text{S}^1\times\text{S}^2)$.

Once the energy surfaces $\Ehl$ are given, the problem is to find how they are partitioned
into regular and irregular types of motion, i.\,e., into invariant subsets of one, two or three
dimensions (isolated periodic orbits, resonant or non-resonant tori, and chaotic regions,
respectively).
The most valuable tool for this kind of studies is the method of Poincar\'e sections. It requires
the identification of a 2-D surface of section $\Phl \subset \Ehl$ which intersects all possible
trajectories. Finding such a global Poincar\'e section is a non-trivial matter.
It is in general not possible to choose a surface which intersects all trajectories
\emph{transversally}~\cite{BDW96}. However, it is possible to find a surface
(or a set of disjoint surfaces) which is \emph{complete} in the sense that every orbit
intersects -- or at least touches -- it repeatedly. A constructive procedure to obtain
such section conditions was given in~\cite{DW95} and will be employed here.

The next question refers to the topology of the surfaces of section, and how the various types
of~$\Phl$ may be adequately represented in two-dimensional plots. This is the main concern of the
present paper. We show that a single connected component of $\Phl$ may be either a sphere~S$^2$,
a torus~T$^2$, or a 2-D manifold M$^2_g$ of genus~$g=2$, 3, or 4. A single kind of graphical
representation applies to all cases: the 1:1 projection to a torus constructed from two copies
of a two-fold punctuated Poisson sphere. We call this the ``PP-torus'' T$^2_2(\ga)$ and propose it
as a convenient universal tool for
investigations of the complex dynamics of rigid bodies. For example, it may be interesting to
vary $(h,l)$ for fixed parameters $(A_1,A_2,A_3)$ and~$(r_1,r_2,r_3)$, in order to obtain a
complete survey of the phase space structure for a given rigid body. For the integrable
Kovalevskaya top such a survey was presented in~\cite{RDW97}, but the new tool can be applied to
any choice of parameters. Alternatively, one might want to vary parameters at fixed $(h,l)$,
and to follow the fate of certain conspicuous features (like isolated periodic orbits, or major
chaotic regions).

The paper is organized as follows. Section~\ref{Sec:Top} recalls how the topology of energy
surfaces is determined from the effective potential on the Poisson sphere. The new Poincar\'e
section is introduced in Section~\ref{Sec:PSS} and compared to a proposal made earlier
in~\cite{GR2004}. Finally, Section~\ref{Sec:Phl} describes the topology of the surfaces of
section~$\Phl$ and their projection first to the Poisson sphere, then to the PP-torus constructed
from it.

\section{Topology of energy surfaces}
\label{Sec:Top}

The heavy rigid body on T$^*$SO(3) is symmetric with respect to rotation
about the axis of gravity. Reduction by this symmetry gives
the Euler-Poisson equations for the motion of a rigid body about a fixed point
as seen in a co-moving frame,
\begin{equation}
\label{ode}
\begin{split}
\dot{\ga} &= \left\lbrace \ga,H \right\rbrace
  = \ga \times \boldsymbol{\omega} \\
\dot{\mathbf{l}} &= \left\lbrace \mathbf{l},H \right\rbrace
= \mathbf{l} \times \boldsymbol{\omega} - m g \ga \times \mathbf{r} \,.
\end{split}
\end{equation}
Here
$\ga$ is the unit vector along the spatial $z$-axis (the axis of gravity),
$\l$ is the angular momentum vector,
$\om = \A^{-1} \l$ the angular velocity vector,
$\A = \text{diag}(A_1,A_2,A_3)$ is the matrix of principal moments of inertia,
and $\r$ is the position of the center of mass in the body.
The connection to Euler's angles with respect to the $z$-axis is
$(\gamma_1,\gamma_2,\gamma_3) = (\sin\psi\sin\vartheta,\cos\psi\sin\vartheta,\cos\vartheta)$;
the angle~$\varphi$
of rotation about the $z$-axis is eliminated by the symmetry reduction.
The remaining angles $(\vartheta,\psi)$ parameterize the Poisson sphere~S$^2$.
The constants $mg$ can be absorbed into $\r$ and will be ignored from now on.
These equations on the reduced phase space
are Hamiltonian with respect to the Poisson bracket
\begin{equation}
\label{EM:equ6}
\left\lbrace F, G \right\rbrace = \langle \nabla_{\ga}F,
\ga \times \nabla_{\mathbf{l}}G\rangle  +
\langle \nabla_{\mathbf{l}}F, \ga \times \nabla_{\ga}G
+ \mathbf{l} \times \nabla_{\mathbf{l}}G\rangle
\end{equation}
and Hamiltonian
\begin{equation} \label{EM:equ2}
H( \ga, \l) =\frac{1}{2} \langle \l , A^{-1} \l \rangle -  \langle \ga, \r \rangle  \,,
\end{equation}
where $\langle \cdot, \cdot \rangle$ denotes the standard Euclidean scalar product in $\R^3$.
The bracket has two
Casimirs\footnote{note that $l$ does not denote the length of $\l$, but the value of
its $z$-component $\langle\ga, \l \rangle$.},
\begin{equation}
\label{EM:equ8}
I(\ga) = \langle \ga,\ga\rangle = 1  \quad \text{and} \quad
L_z(\ga, \l) = \langle\ga, \l \rangle = l  \,.
\end{equation}
Fixing the Casimirs to the values $I(\ga) = 1$ and $L_z(\ga, \l) = l$
defines the reduced phase space T$^*_l$S$^2$.
Fixing, in addition, the energy $H= h$ defines the reduced energy surface $\Ehl$.
The effective (or amended) potential on S$^2$ is, see e.\,g.~\cite{RDW97},
\begin{equation}
\label{EM:equ10}
U_l(\ga) = \frac{l^2}{2\langle
\ga,A \ga\rangle } - \langle \ga,\mathbf{r}\rangle \,.
\end{equation}
The term proportional to $l^2$
(the kinetic energy of rotation about the axis of gravity)
contains the moment of inertia $\langle \ga, A \ga \rangle$.
\wegPeter{After reduction by the rotational symmetry about the $z$-axis
the overall kinetic energy will always at least as big as this term.}  
The accessible region on S$^2$ for fixed energy $h$ and angular momentum $l$ is
\begin{equation}
\label{EM:equ9}
\Uhl = \left\lbrace \ga :  U_l(\ga) \leq h \right\rbrace
\end{equation}
The energy-Casimir map is given by
\begin{equation}
\label{EMM:equ1}
\begin{aligned}
\F : \R^6 &\longrightarrow \R^3 (h,l,1) \\
(\ga, \l) & \longmapsto \bigl((H(\ga, \l), L_z(\ga, \l), I(\ga)\bigr)
\end{aligned}
\end{equation}
Its critical points $(\ga, \l):\,\text{rank}\,\text{D} \F < 3$,
are the relative equilibria\footnote{see Arnold \cite{Arnold78}, Appendix 5C.},
and their images under $\F$ are the critical values.
Since the last component of the energy-Casimir map is constant,
it suffices to consider the $(h,l)$ energy-momentum plane as the image.
The energy surface  $\Ehl$ is the preimage of the point $(h,l)$.
The set of critical values is called the bifurcation diagram.
For critical values the energy surface is not a smooth manifold, and the topology
in general changes upon crossing critical values.

\begin{figure}
\begin{center}
\includegraphics[width=0.45\textwidth]{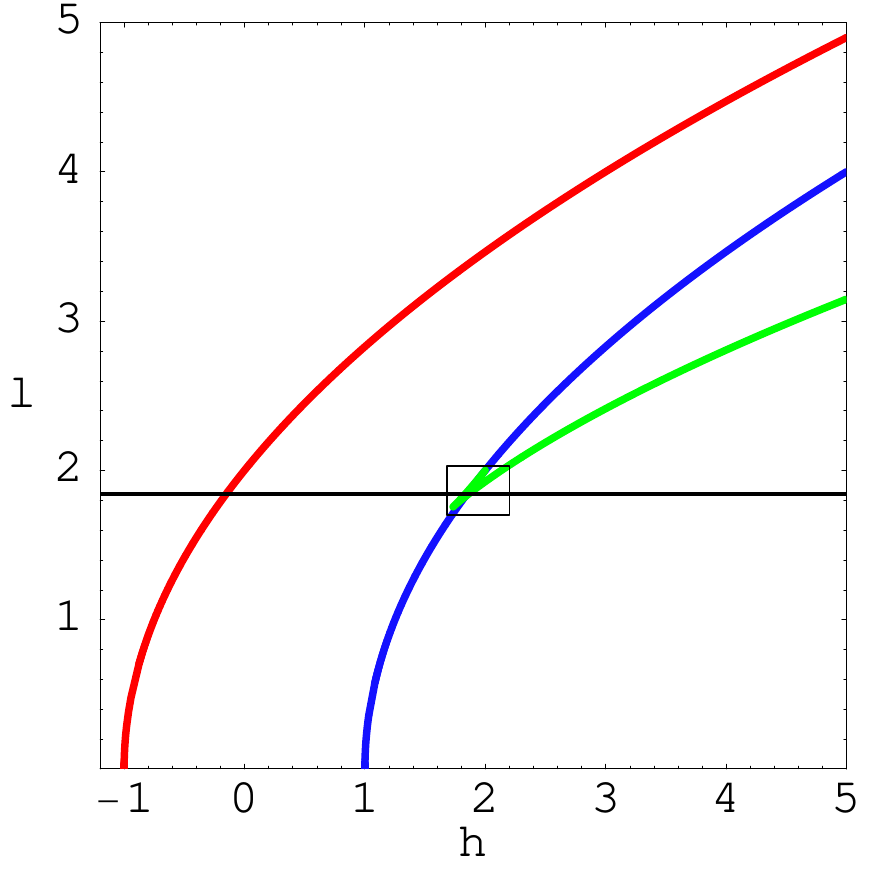}
\hspace{1cm}
\includegraphics[width=0.45\textwidth]{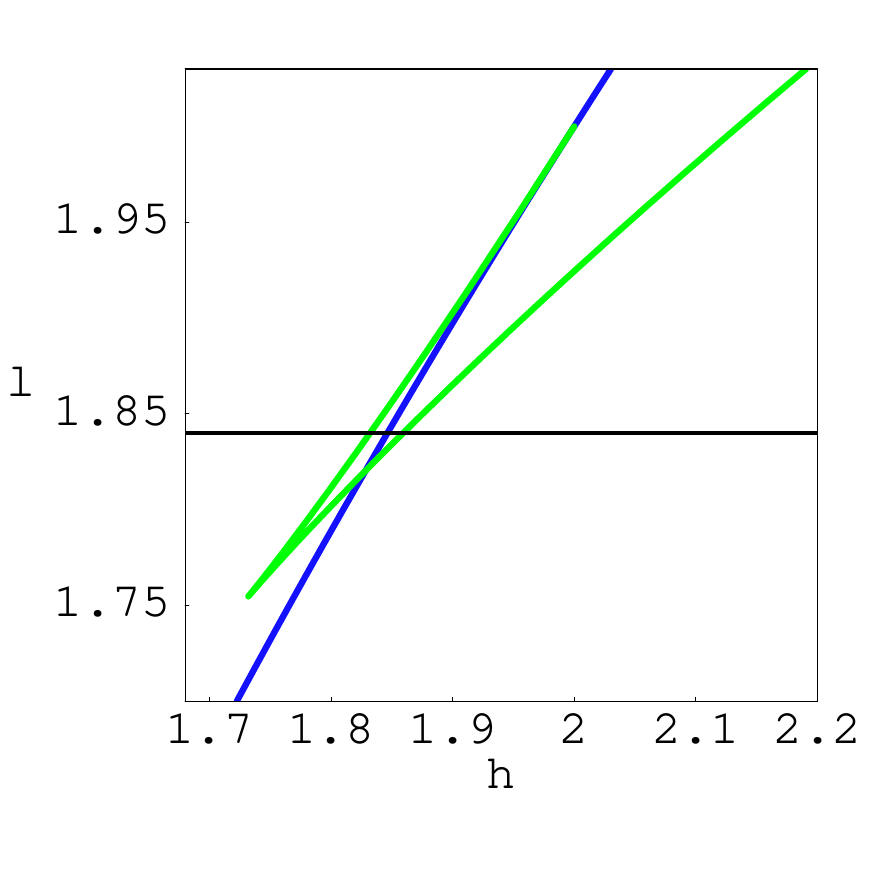}
\end{center}
\caption{\label{Fig:BifDiagKowa}\small The bifurcation diagram showing
critical values $(h,l)$ of the energy-Casimir map $\F$ for
the Kovalevskaya case $\A={\rm diag}(2,2,1)$, $\r=(1,0,0)$.
Right: Magnification of the rectangle shown on the left.
}
\end{figure}

Instead of computing the rank of~$\F$, the equilibria of~\eqref{ode} can be computed directly.
The relation between the two approaches is that the gradients of the Casimirs
are in the kernel of the Poisson structure, and if rank\,D$\F$ is lower than~3, the gradient of the
Hamiltonian is zero or a linear combination of the gradients of the Casimirs.
A third method to obtain the relative equilibria is to compute the critical points
of the effective potential~$U_l$. The energy surface can be viewed as a singular circle bundle
over the accessible region $\Uhl$, see e.\,g.~\cite{BDW96}. At critical points of~$U_l$ the
topology of $\Uhl$ changes, inducing a change in the topology of
the energy surface.
Thus a bifurcation diagram as shown in Fig.~\ref{Fig:BifDiagKowa} can either be read as
a statement about the existence of relative equilibria at the critical values shown,
or as a statement about the topology of the reduced energy surface for the non-critical values.

For example~\cite{RDW97}, along the line $l=1.84$ in Fig.~\ref{Fig:BifDiagKowa}, the topologies
of the
accessible regions $\Uhl$ are, from left to right,
empty set, disk, disk with two holes, disk with one hole, and all of S$^2$.
The transitions take place when crossing the lines of critical values
in the bifurcation diagram.
The corresponding topologies of the energy surface $\E^3_{h,l}$
are determined by the topology of the accessible region, see~\cite{BDW96}.
They are
$\emptyset$, S$^3$, (S$^1 \times $S$^2$) $\#$ (S$^1 \times $S$^2$), S$^1 \times $S$^2$, $\R$P$^3$,
again from left to right.
More general cases of the bifurcation diagram have been discussed
in~\cite{Katok72, Tatarinov74}, see also~\cite{GR2004}.

It should be remembered that of the three Euler angles, only $\vartheta(t)$ and $\psi(t)$ appear
in the Euler-Poisson equations;
the angle~$\varphi(t)$ must be determined by integration of~\cite{RDW97}
\begin{equation}
 \dot{\varphi} = \frac{l_1\gamma_1 + l_2\gamma_2}{2(\gamma_1^2 + \gamma_2^2)}.
\end{equation}
Hence, relative equilibria $\ga = \text{const}$, $\l = \text{const}$
of the symmetry reduced system correspond to periodic motion of the
full system.

\section{Poincar\'e surface of section}
\label{Sec:PSS}

For systems with two degrees of freedom and 3-dimensional compact energy surfaces, the
relevant information about phase space structure (stability of periodic orbits,
relative extent and entanglement of regular and chaotic motion) is contained
in the 2-dimensional maps induced by the motion on suitably chosen Poincar\'e
surfaces of section. This has become a standard tool from celestial mechanics
to molecular dynamics, yet in the study of rigid bodies there have so far
been only few applications~\cite{Dullin94b, DJR94, RichNPCS07}. However, as the Euler-Poisson
equations
describe a system with effectively two degrees of freedom (at given~$l$),
Poincar\'e sections are the ideal method to study the complexity of
rigid body dynamics. In the following, we propose a variant which is at the
same time general and easy to implement. We begin with a few general considerations.

\subsection{General features of Poincar\'e sections}

It would be nice to identify a two-dimensional submanifold $\Phl$ of the energy surface
$\Ehl$ in such a way that every orbit meets it repeatedly and transversally. As was shown
in~\cite{DW95}, \emph{complete} sections can indeed be defined, i.\,e.\ sections which
capture every single trajectory. However,
globally transverse sections do not exist in general, for topological reasons~\cite{BDW96}.
Namely, when the surface of section $\Phl$ divides $\Ehl$ into an ``inner'' and an ``outer'' part,
trajectories that ``come in'' have to ``go out'' again, so that part of $\Phl$ is traversed by
ingoing,
another part by outgoing orbits. The boundary between the two is then a subset of $\Phl$
where the trajectories are tangent.

A complete surface of section
$\Phl$ can be obtained with a recipe from~\cite{DW95}:
given any smooth function~$W$ which maps the energy surface into a bounded set,
define $S := \dot W$ and take $S(\ga, \l) = 0$ as section condition:
\begin{equation}
\label{PSS:equ1}
\Phl := \left\{ (\ga,\l)
\in
\Ehl :   S(\ga, \l) = 0
\right\} \,.
\end{equation}
$\Phl$ is a smooth 2-dimensional manifold except when the map
\begin{equation}
\label{EMM:equ2}
\begin{aligned}
P : \Ehl &\longrightarrow \R^4 (h,l,1,0) \\
\left( \ga, \l  \right) & \longmapsto (H(\ga, \l), L_z(\ga, \l), I(\ga), S(\ga,l))
\end{aligned}
\end{equation}
has a critical point indicated by rank\,D$P(H, L_z, I, S) < 4$.
Clearly the rank of P drops when the energy surface is not a smooth manifold,
but there may be additional singular values.

The tangent set, if it exists, is given by those points on $\Phl$ for
which $\dot S = 0$. Consider for example the Hamiltonian
$H = \frac{1}{2}p_1^2 + \frac{1}{2}p_2^2 + V(q_1,q_2)$ and the section condition
$S = q_2 = 0$.\footnote{This section condition is not of the form $S = \dot{W}$, hence it need not
be complete;
there may be orbits which never cross the line $q_2=0$ in the $(q_1,q_2)$-plane.}
The surface of section is then given by ${\cal{P}}^2_h = \{ (q_1,p_1,p_2):
\frac{1}{2}p_1^2 + \frac{1}{2}p_2^2 + V(q_1,0) = h\}$. Its usual representation is in terms of its
projection to the $(q_1,p_1)$-plane which is a 2:1-map. To make it unique, only the
part $\dot{q}_2 > 0$ is considered; the part $\dot{q}_2 < 0$ is ignored because
the two parts are related by time reversal and contain no independent information.
The line of tangency $\dot{S} = \dot{q}_2 = p_2 = 0$ lies in the projection plane and is in fact
the boundary of the energetically accessible region $\{ (q_1,p_1):
\frac{1}{2}p_1^2 + V(q_1,0) \leq h\}$.

\begin{figure}
\begin{center}
\includegraphics[width=0.45\textwidth]{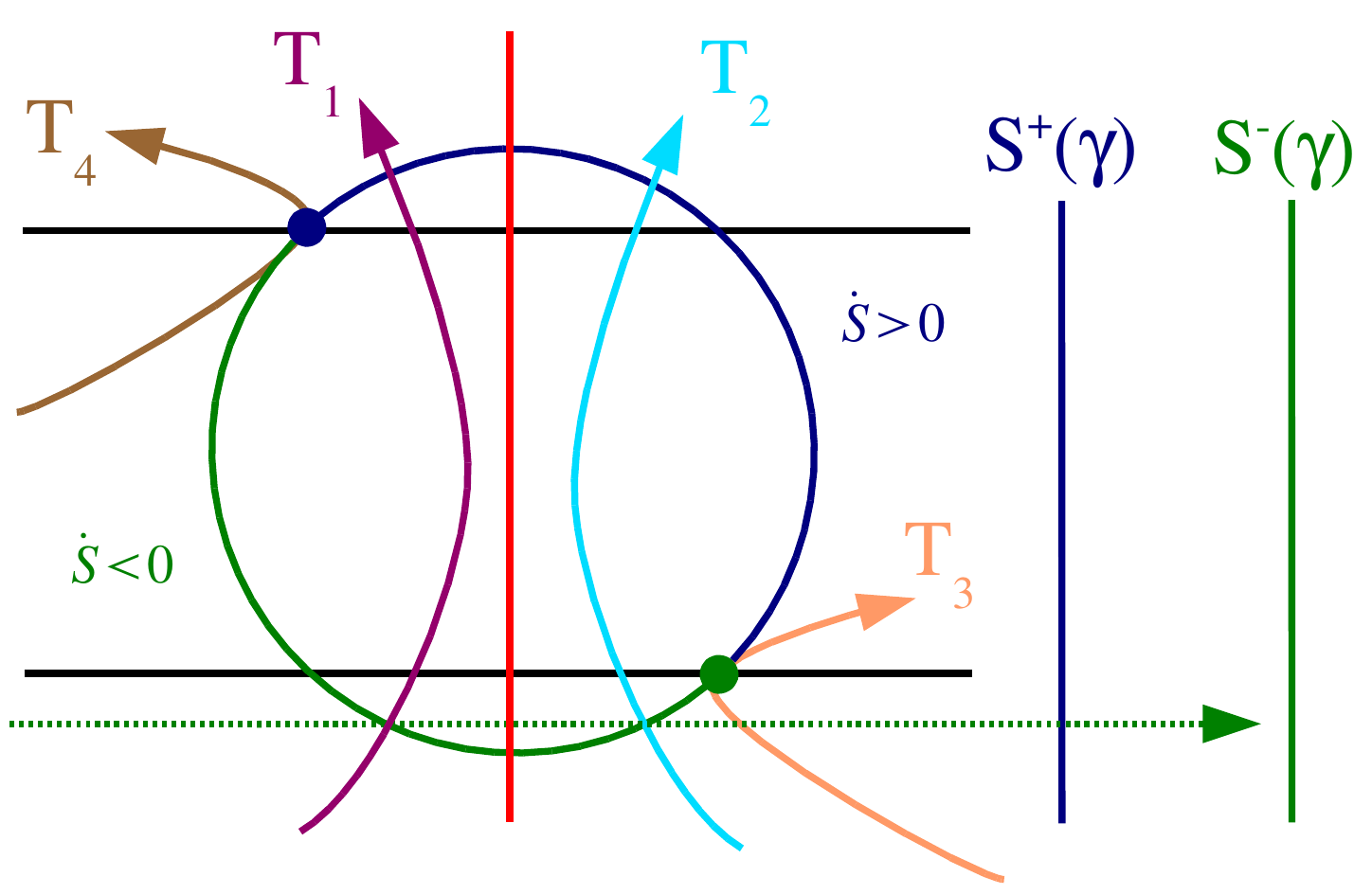}
\includegraphics[width=0.45\textwidth]{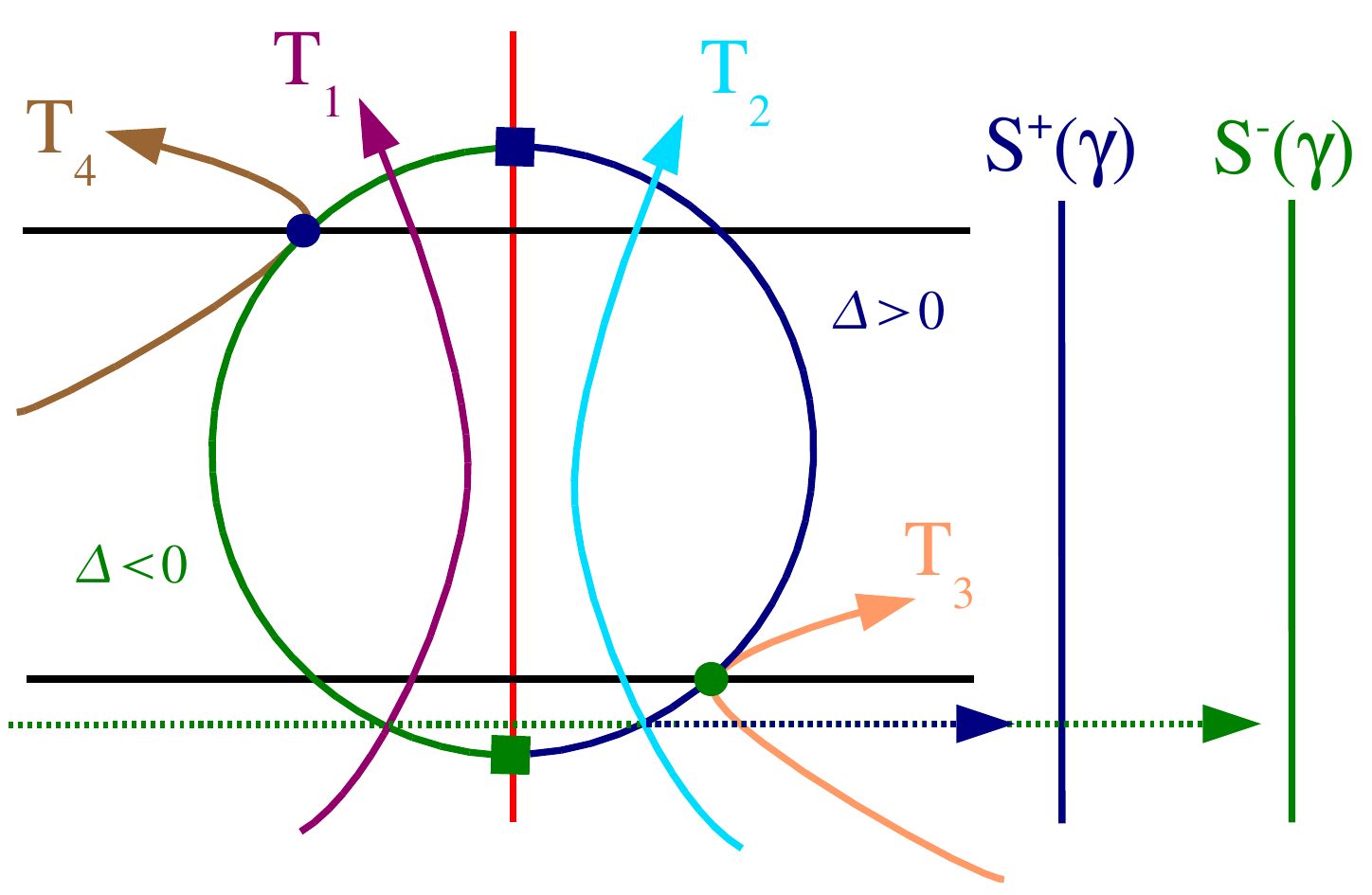}
\end{center}
\caption{\label{Fig:schema12} \small Schematic presentation of the Poincar\'e surface
of section~$\Phl$ (circle) with four trajectories two of which are tangent in the thick dots,
and its projections onto two copies of the sphere S$^2(\ga)$ (vertical bars).
Left: S$^\pm$ defined by the sign of~$\dot S$ (projections not invertible).
Right: S$^\pm$ defined by the sign of~$\Delta$.
}
\end{figure}

The situation is similar but somewhat more complicated in the case
of rigid body dynamics. Assume we have a section condition $S(\ga,\l)=0$. The
corresponding surface $\Phl \subset \Ehl$ ``lives'' in the 6-dimensional $(\ga,\l)$-space,
$\ga \in \text{S}^2$ and $\l \in \R^3$. A convenient projection, as we shall see, maps it
onto the Poisson sphere, $\pi: \Phl \to \text{S}^2(\ga)$. Here it is
not generally true that the boundary of the projection coincides with the points
of tangency. The situation is rather as shown schematically in the left part
of~Fig.~\ref{Fig:schema12}.
There $\Phl$ corresponds to the circle, and the bars to its right are meant to represent two copies
of S$^2(\ga)$. The four trajectories illustrate where $\Phl$ intersects incoming and
outgoing orbits ($\dot S < 0$ and $\dot S > 0$, respectively). In this sketch,
the line of tangency reduces to two points which may lie anywhere on the circle. Then, if
S$^+(\ga)$ were chosen to carry the projection of the part where $\dot S > 0$, and likewise
S$^-(\ga)$ for $\dot S < 0$, these maps would obviously not be 1:1. In this setting,
$\ga$ could not serve as a local coordinate on~$\Phl$.

On the other hand, consider the right part of Fig.~\ref{Fig:schema12}. There the circle representing
$\Phl$
is divided by the two solid squares in such a way that the two projections to S$^\pm(\ga)$ are
indeed 1:1.
A drawback of this welcome feature is that both projections contain incoming and outgoing
intersections;
the points of tangency no longer form the boundaries of the projections. But this turns out not to
be
serious. The important point is that the projections S$^\pm(\ga)$ provide local coordinates on the
two halves
of~$\Phl$. But how is this schematic picture to be implemented? It must be possible to uniquely
determine
the momenta~$\l$ on~$\Phl$ from the coordinates~$\ga$. With $\Phl$ defined by $H=h$, $L_z=l$, and
$S=0$, the
implicit function theorem guarantees that this can be done unless
\begin{equation}
\label{IFTdet}
 \Delta(\ga,\l) := \text{det} \frac{\partial\ }{\partial\l}(H,L_z,S) = 0.
\end{equation}
The relevant division of $\Phl$ into two parts is therefore given by the sets
$\Pp \subset \Phl: \Delta > 0$ and $\Pm \subset \Phl: \Delta < 0$ which project to S$^+(\ga)$ and
S$^-(\ga)$ respectively. The condition~\eqref{IFTdet} contains the definition of the dividing line
on~$\Phl$.
It will be seen in~\eqref{eq:Delta} that there may exist other lines in~$\Phl$ where $\Delta=0$,
but these project only to points in S$^2(\ga)$.

\begin{figure}
\begin{center}
\includegraphics[width=0.45\textwidth]{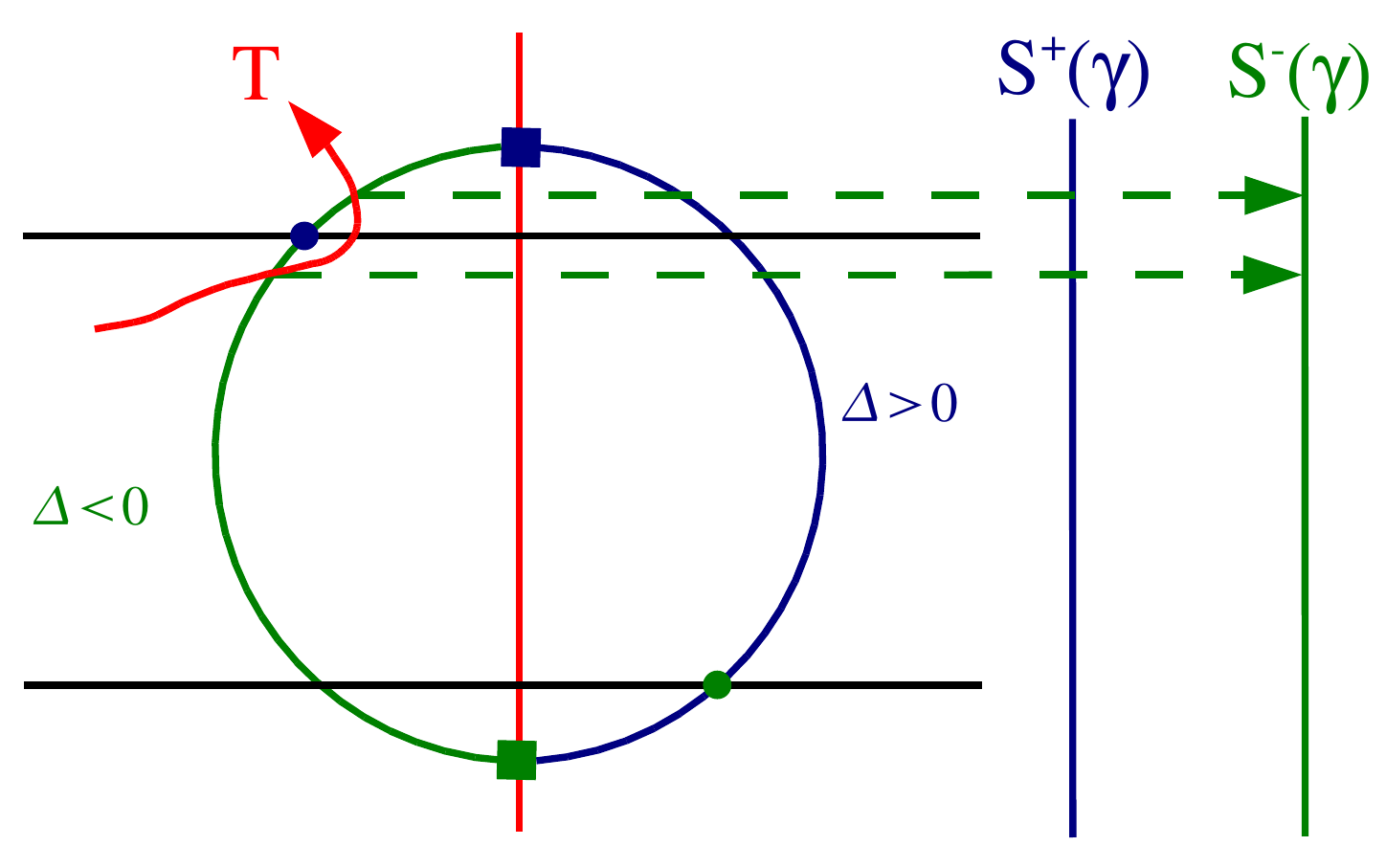}
\end{center}
\caption{\label{Fig:schema3} \small Only both copies of S$^2(\ga)$ allow a complete characterization
of the phase space structure since intersection points can be projected to one copy only.}
\end{figure}

Note that even when the system is symmetric under time reversal (which is not the case here,
at fixed~$l\neq 0$), it may not be sufficient to consider only one of the two projections.
As can be seen in Fig.~\ref{Fig:schema3}, orbits with intersection points close to tangency tend to
have incoming and outgoing intersections with the same sign of~$\Delta$; hence, in order to capture
all
orbits, both projections are needed.

\subsection{The section condition}

The first general section condition of the type $S = \dot{W}$ was proposed in~\cite{GR2004}
with $W = \langle \l, \l \rangle$. It was motivated by a study of the projections of the
energy surfaces $\Ehl$ to $\l$-space. There the preimages of the envelopes of these projections
could be
characterized by $S=0$, or $\langle \l, \ga \times \r \rangle = 0 $,
i.\,e.\ in terms of local extrema of the values of the total angular momentum.
Clearly this section condition is the adequate choice when the envelopes in $\l$-space are used to
represent the Poincar\'e surfaces of section; a number of beautiful examples have been worked out in
this setting (Gashenenko, private communication). However, the complex topology of the corresponding
surfaces~$\Phl$ (being manifolds of genus up to five and even higher) makes it hard to study them.
Furthermore, the $\l$-envelopes are different for each point $(h,l)$, and difficult to parameterize,
hence they do not lend themselves for extended investigations. An obvious way out of this last
difficulty
is to project the surfaces $\Phl$, defined by $W = \langle \l, \l \rangle$, to the Poisson
sphere~S$^2(\ga)$.
However, it turns out~\cite{GR2004} that this produces artifacts in the projections which have no
relevance in~$\Phl$ itself:
the condition $W = \langle \l, \l \rangle$ is well adapted for $\l$-space, not for $\ga$-space.

Therefore we propose another choice:
\begin{equation}
 \label{eq:PSecCond}
  W = \langle \ga,\r \rangle  \qquad \Rightarrow \qquad S(\ga,\l) = \frac{\dee W }{\dee t} =
      \langle A^{-1}\l , \r \times \ga \rangle = 0 \, .
\end{equation}
This is motivated by the behavior of the Lagrange top where $\r$ lies on the body axis so that
$\langle \ga,\r \rangle \propto \cos\vartheta$. The angle~$\vartheta$
oscillates between $\vartheta_{\rm min}$ and $\vartheta_{\rm max}$ which
makes $\dot \vartheta = 0$ a natural section condition.
In addition, when successive points $(\vartheta_n,\psi_n)$ ($n=0, 1, 2, ...$) of the corresponding
Poincar\'e map are plotted on S$^2(\ga)$, a natural winding number may be read off in terms of
the increments of $\psi$.
The section condition~\eqref{eq:PSecCond}
is an obvious generalization; it picks out
extrema of the projection of the center of mass~$\r$ to the vertical direction.
As we will show, the resulting surfaces of section $\Phl$ have genus only up to 4,
and they have fewer bifurcations than those studied in~\cite{GR2004}.
Moreover, the projections  $\pi(\Phl)$
onto the Poisson sphere coincide with the accessible region $\Uhl$,
and the condition $\Delta(\ga,\l)=0$ defines a line which projects to the boundary~$\partial\Uhl$.
Finally, this projection is simple because generically it has
exactly two preimages; the exceptions are $\ga \in \partial\Uhl$ which have only one preimage, and
$\ga\parallel\r$ where the preimage is a circle.

To prove these statements, we simultaneously solve the equations $H=h$, $L=l$ (which together define
the energy surface~$\Ehl$) and $S=0$; i.\,e., given a point $\ga$ on the Poisson sphere, we determine
the set of corresponding values $\l$ such that $(\ga,\l)\in \Phl$. The explicit calculation is given
in the Appendix. There we excluded the possibility that $\ga$ and $\r$ are collinear,
$\r\times\ga = 0$,
so let us start here with this case: $\ga = \pm \r/r =: \hat{\r}$.
The condition $S=0$ is then identically fulfilled and gives no restriction
on~$\l$. The condition $L_z = \langle \l,\ga\rangle = l$ defines a plane in $\l$-space, and $H=h$
the ellipsoid
$\langle \l, A^{-1}\l\rangle = 2(h+\langle\r,\ga\rangle) = 2(h \pm r)$. Three cases are possible:
\begin{itemize}
 \item[(i)] The intersection of plane and ellipsoid is a topological circle.
 \item[(ii)] They do not intersect at all.
 \item[(iii)] The plane is tangent to the ellipsoid in a point $\l^\ast$.
\end{itemize}
In the latter case the normal to the ellipsoid must be collinear with $\ga$,
i.\,e.\ $A^{-1}\l^\ast = \xi\ga$ or
$\l^\ast = \xi A\ga$. From $\langle \l^\ast,\ga\rangle = l$ we obtain $\xi = l/\langle
\ga,A\ga\rangle$, hence the
energy equation becomes
\begin{equation}
 \label{eq:collinear}
 \frac{l^2}{\langle \ga,A\ga\rangle} = 2(h\pm r) \qquad \Leftrightarrow \qquad h = U_l(\ga) =
 U_l(\pm \hat{\r})\, .
\end{equation}
So in this case, $\ga = \pm \hat{\r}$ lies on the boundary of the accessible region, $\ga \in \partial\Uhl$.
In case (i),
$\ga$ lies inside, in case (ii) outside. This will be relevant for the topology of~$\Phl$,
see next section.

Let us now consider $\ga$ which are not collinear with $\r$.
We show in the Appendix that given $\ga$, the preimages in
$\Phl$ are the points $(\ga,\l)$ with
\begin{equation}
 \label{eq:lofgamma}
 \l = l\,\frac{A\ga}{\langle \ga, A\ga \rangle} \pm \v\,\sqrt{\frac{2(h-U_l(\ga))}{\langle
 \v,A^{-1}\v \rangle }}\, ,
\end{equation}
where $\v = \bigl( A^{-1}(\r\times\ga) \bigr)\times \ga$. This tells us that for all $\ga$ with
$U_l(\ga) < h$, or $\ga$ from the interior of $\Uhl$, there are exactly two preimages in $\Phl$.
For $\ga$
on the boundary, $\ga \in \partial \Phl$, the preimage is unique. No real $\l$ exists for $\ga$
outside $\Uhl$.

\weglassen{
A somewhat trivial special case is $h + \langle \ga,\r \rangle = 0$, corresponding to a circle
on the Poisson sphere with center $\r$. With~\eqref{EM:equ10} we see that this
can only happen for $l=0$, and~\eqref{eq:lofgamma} gives $(\ga,\l) = (\ga,0) \in \Phl$. Again,
this $\ga$ is from the boundary of ${\cal{U}}_{h,0}$, and for $\ga$ such that $U_0(\ga) < h$,
\eqref{eq:lofgamma} defines two values of~$\l$.
}

Computing the determinant $\Delta(\ga,\l)$ according to Eq.~\eqref{IFTdet}, we find
\begin{equation}
 \label{eq:Delta}
 \Delta(\ga,\l) = \langle A^{-1}\l\times\ga, A^{-1}(\ga\times\r)\rangle \, .
\end{equation}
This is zero for $\ga \in \partial \Uhl$, where $A^{-1}\l\times\ga=0$, corroborating the assertions
made in connection with the right panel of Fig.~\ref{Fig:schema12}: the parts of $\Phl$
where $\Delta >0$
and $\Delta < 0$ each project to the entire accessible region $\Uhl \subseteq \text{S}^2(\ga)$.
(Note that $\Uhl$, hence also $\Ehl$ and $\Phl$, may consist of several disconnected components.)
The other possibility for $\Delta$
to become zero, $\ga\times\r = 0$, is of another kind; its projection to S$^2(\ga)$ gives at most
the two points $\ga = \pm \hat{\r}$. If they are inside $\Uhl$, their preimages in $\Phl$
are circles, see case (i) above Eq.~\eqref{eq:collinear}.
The consequences are discussed in the next section.

The condition $\dot{S} = 0$ for tangency of trajectories with $\Phl$ is obtained
from~\eqref{eq:PSecCond}
by differentiation:
\begin{equation}
  \dot{S} = \langle A^{-1}(\l\times A^{-1}\l - \ga\times\r),\r\times\ga\rangle +
            \langle A^{-1}\l,\r\times (\ga\times A^{-1}\l)\rangle = 0 \, .
\end{equation}
This must be considered together with~\eqref{eq:lofgamma}. We do not attempt to derive an explicit
expression for the projection of
these lines to the Poisson sphere, but they will be shown in the graphical representations of the
following sections.

\section{Topology of $\Phl$ and its projections}
 \label{Sec:Phl}

\begin{figure}[H]
\begin{minipage}[t]{\textwidth}
\begin{multicols}{2}
\begin{center}
\includegraphics[width=0.4\textwidth]{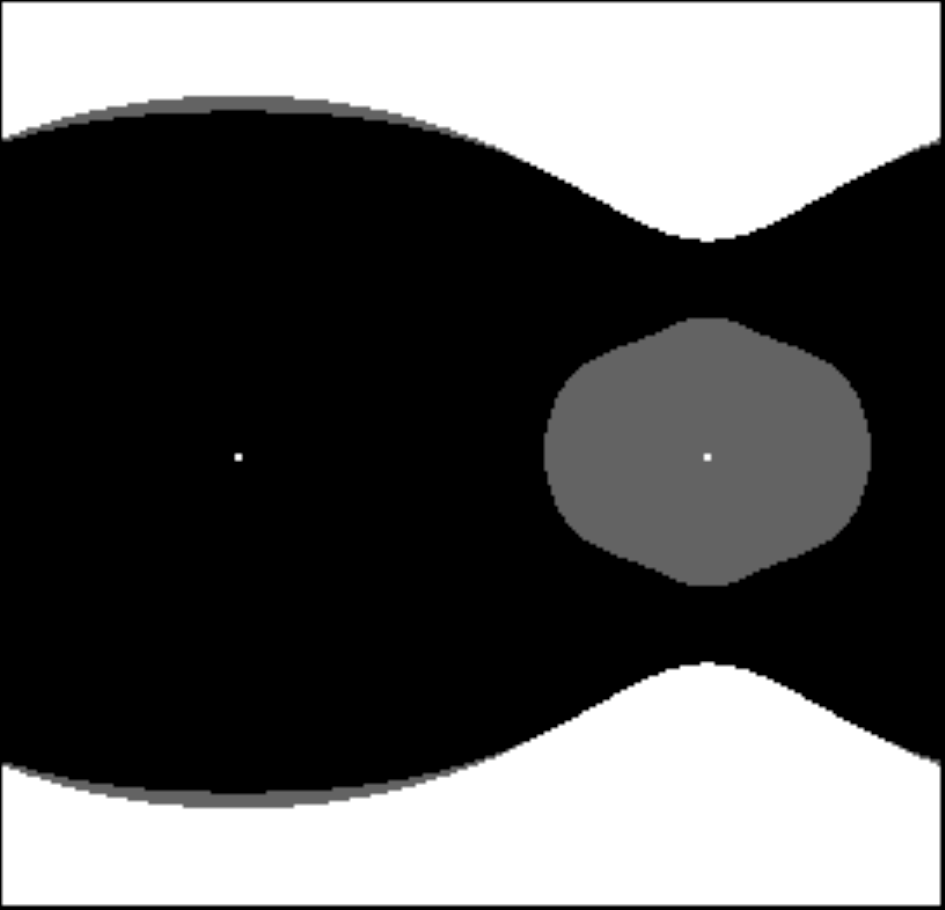}
\end{center}
\newpage
\begin{center}
\includegraphics[width=0.4\textwidth]{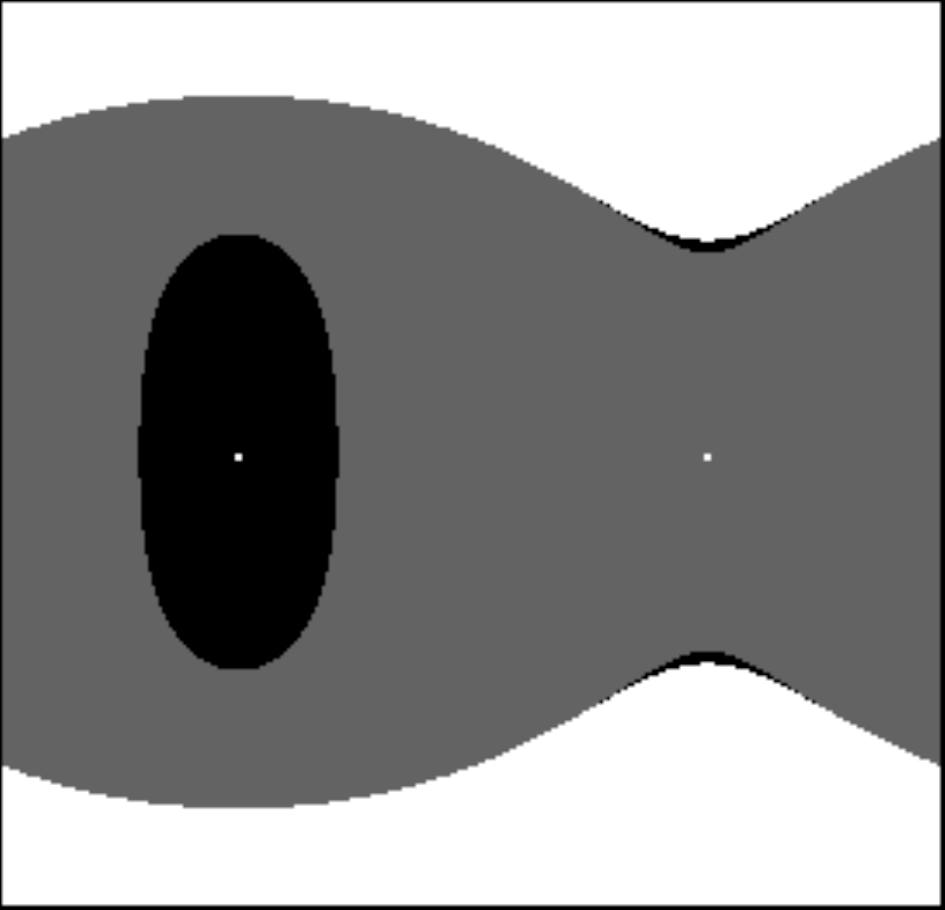}
\end{center}
\end{multicols}
\caption{\label{Fig:projpm} \small Projection of $\Phl$ onto the
Poisson spheres S$^-(\ga)$ (left) and S$^+(\ga)$
(right). On the abscissa $\psi$ goes from $0$ to $2 \pi$, the
ordinate $\vartheta$ varies from $0$ (top) to $\pi$ (bottom). \mbox{$A=(2,2,1)$},
$\mathbf{r}=(1,0,0),
h=3.4, l=2.87$.}
\end{minipage} \\[1ex]
\end{figure}

Fig.~\ref{Fig:projpm} shows a typical projection of~$\Phl$ to two copies
S$^-(\ga)$ (left) and S$^+(\ga)$ (right) of the Poisson sphere,
for $\A=(2,2,1)$ and $\r(1,0,0)$.
The coordinates of the two panels are $0\leq \psi < 2\pi$ on the abscissa and
$0 \leq \vartheta \leq \pi$ on the ordinate. The white regions at small and large $\vartheta$ are
inaccessible with the given values of $h=3.4$ and $l=2.87$. The grey and black parts together are
the accessible region~$\Uhl$ which has the topology of an annulus.
In the black parts the trajectories are incoming, $\dot S < 0$, in the grey region they are outgoing,
$\dot S > 0$; the boundary between these regions is the set of tangency.
If we were to record only incoming trajectories, only the black parts needed to be considered, but
note it appears on both copies~S$^\mp(\ga)$ .

\begin{figure}[H]
\begin{center}
\includegraphics[width=0.5\textwidth]{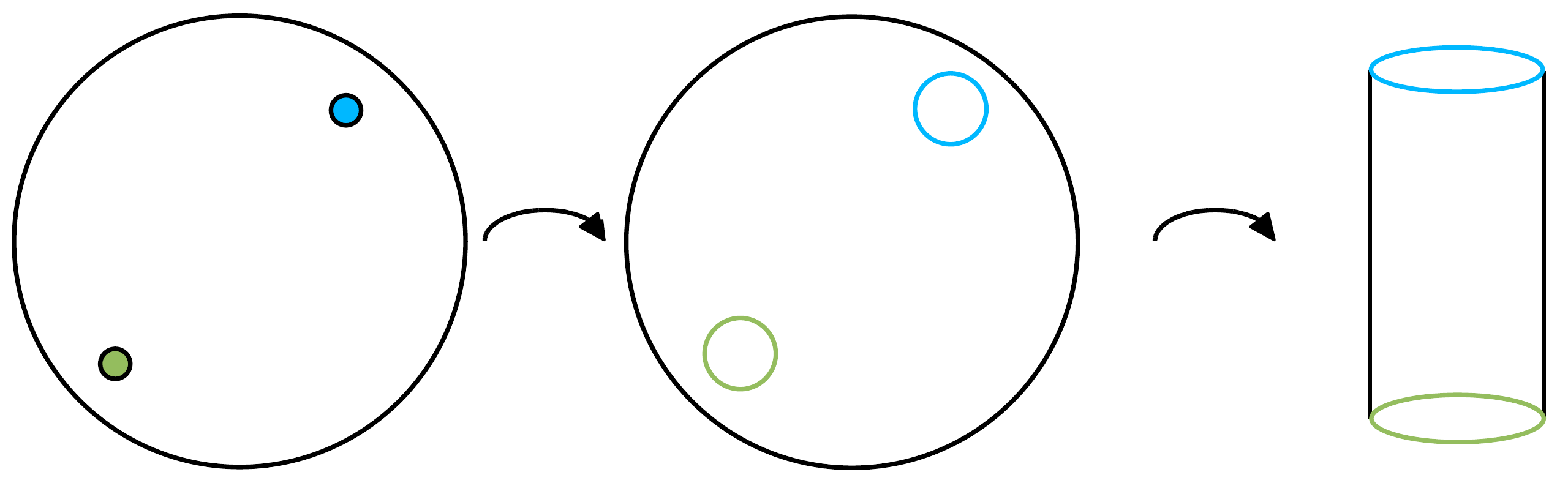}
\end{center}
\caption{\label{Fig:cylinder} \small For a topologically more adequate representation of the
preimages
in $\Phl$ of the poles $\ga = \pm \hat{\mathbf{r}}$, these poles are replaced by
``polar circles'' which turns the Poisson sphere into a cylinder.
}
\end{figure}

The white dots at $(\vartheta, \psi)=(\pi/2,\pi/2)$ are the ``north pole''
$\ga = \hat\r$, the white dots at $(\pi/2,3\pi/2)$ are the ``south pole''
$\ga =  -\hat\r$. We know from the previous section that these points are special. When they
are physically accessible as they are here, then their preimage in~$\Phl$ is a circle (instead of
a point as for all other~$\ga$). This suggests to represent $\Phl$ not in projection to a sphere
but rather to a cylinder: punctuate S$^2(\ga)$ at the poles and insert ``polar circles'' there, as
illustrated schematically in Fig.~\ref{Fig:cylinder}. Instead of the two spheres S$^\mp$ we then
have two cylinders for the projections of $\Pm$ and $\Pp$. But notice that these two cylinders share
the same circles at the poles, hence together they form a torus~T$^2$ which we shall call the
``PP-torus'' T$^2_2(\ga)$, the two P's referring to Poisson and Poincar\'e, or to the two copies of the
punctuated Poisson sphere, and the subscript~2 indicating that each~$\ga$ is represented twice.
We denote its two halves
by T$^-$, T$^+$, and remark that each carries one copy of $\Uhl$. $\Pm$ and $\Pp$,
respectively, project 1:1 to interior points of the two~$\Uhl$, and their boundary, defined by
$\Delta(\ga,\l)=0$, projects to $\partial\Uhl$. Thus, if we identify the two copies of
$\partial\Uhl$ on T$^-$ and T$^+$, we obtain a 1:1 map $\Phl \to \text{T}^2_2(\ga)$
which preserves the topology.

\begin{figure}[H]
\begin{minipage}[t]{\textwidth}
\begin{multicols}{2}
\begin{center}
\includegraphics[width=0.4\textwidth]{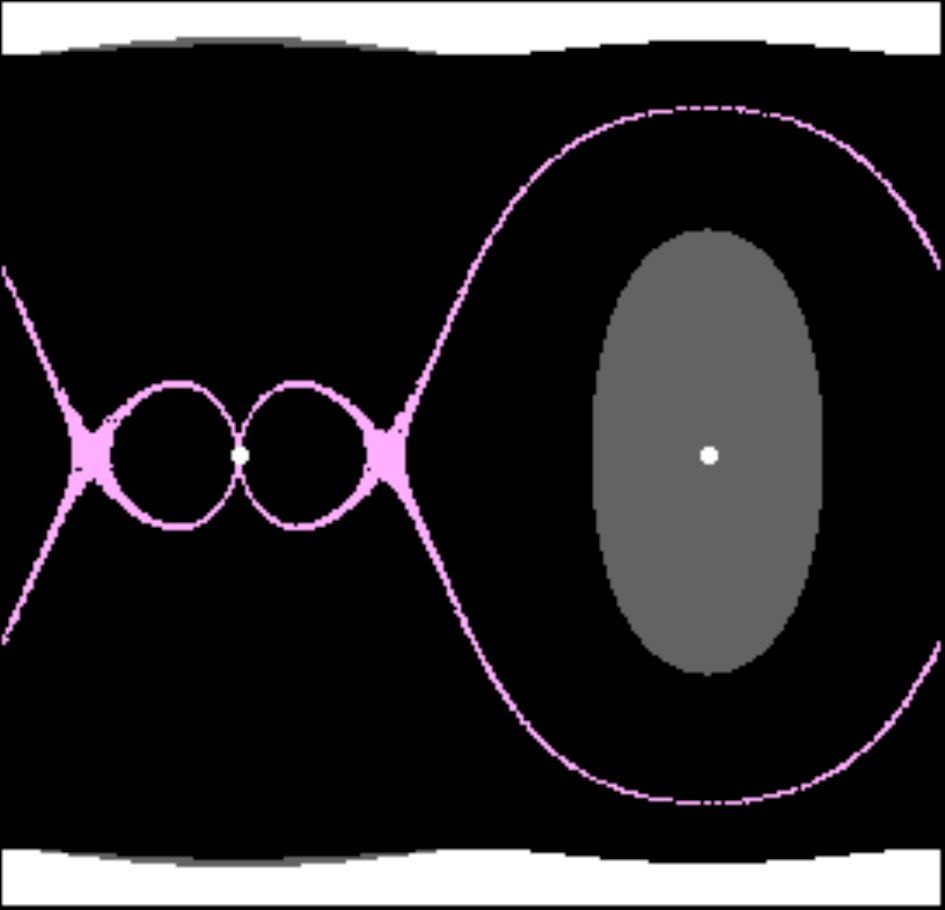}
\end{center}
\newpage
\begin{center}
\includegraphics[width=0.4\textwidth]{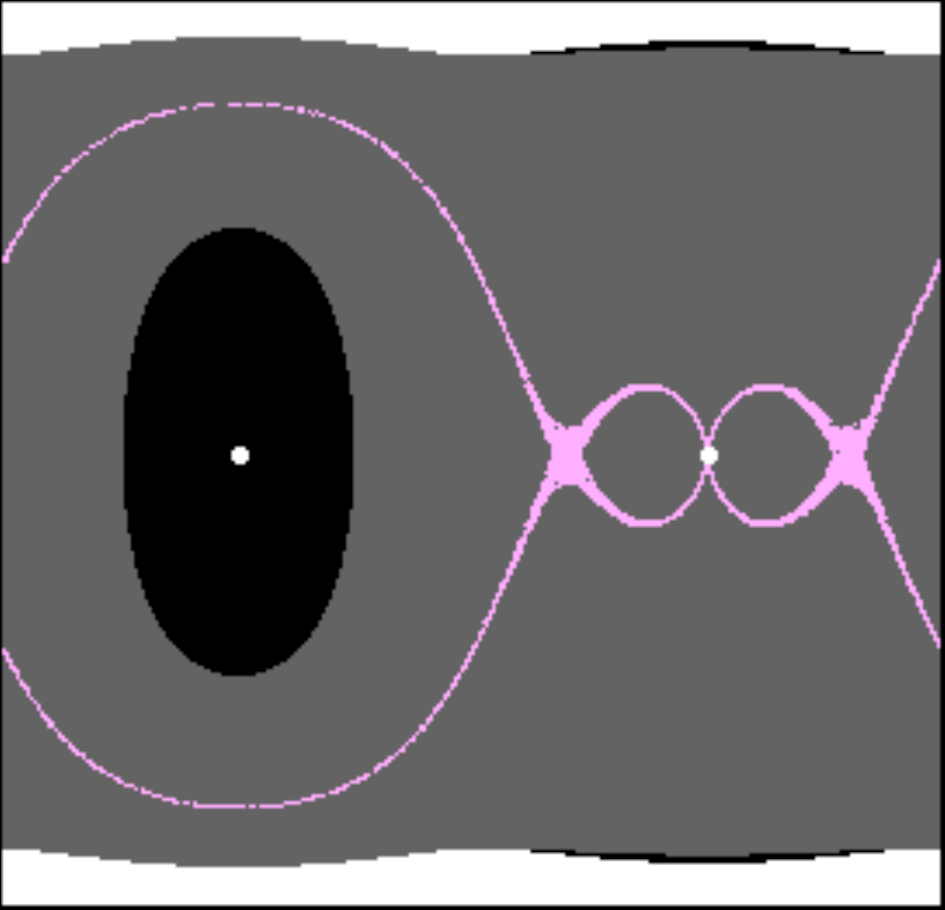}
\end{center}
\end{multicols}
\begin{multicols}{2}
\begin{center}
\includegraphics[width=0.4\textwidth]{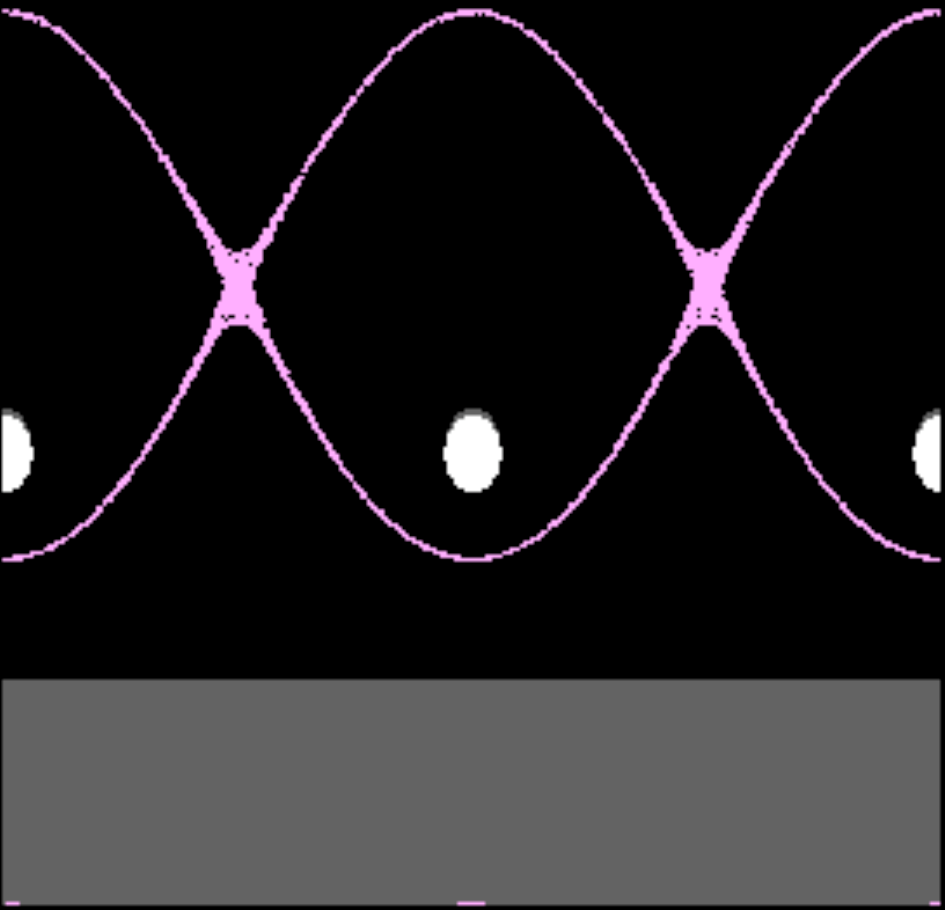}
\end{center}
\newpage
\begin{center}
\includegraphics[width=0.4\textwidth]{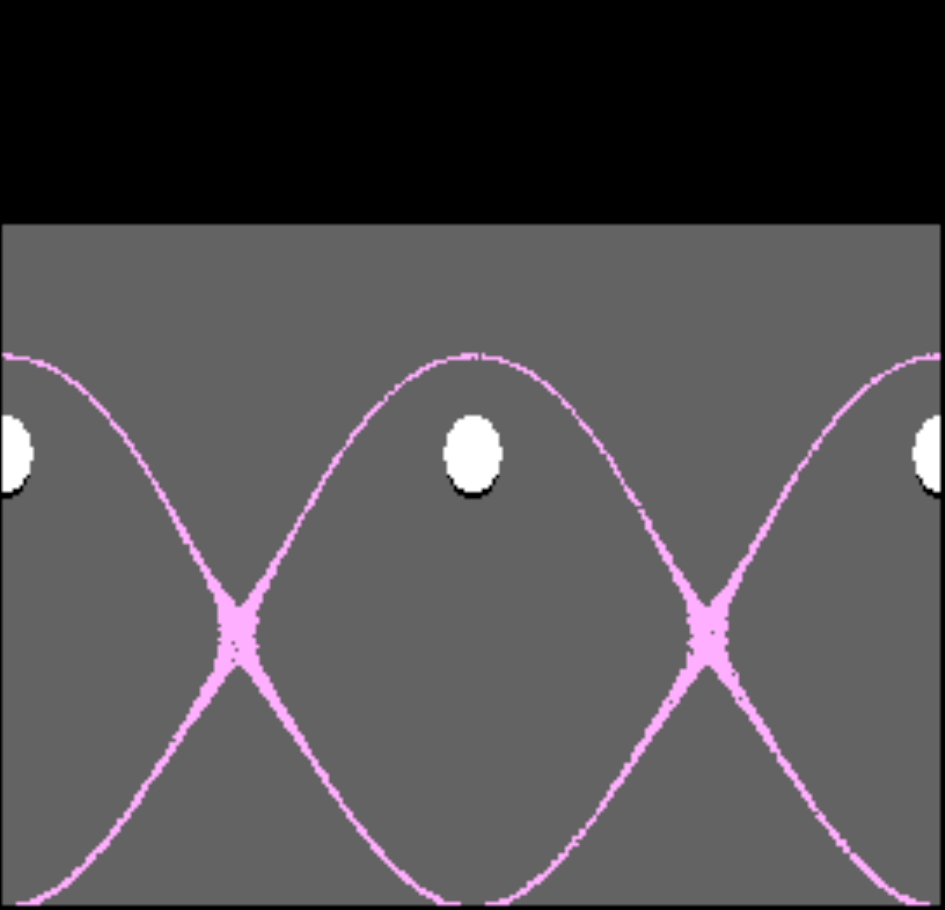}
\end{center}
\end{multicols}
\caption{\label{Fig:Projections12} \small Comparison of the projection of $\Phl$ onto two copies
S$^\mp$ of the Poisson sphere S$^2(\vartheta,\psi)$ (top) and onto the two halves T$^\mp$
of the torus T$^2_2(\vartheta',\psi')$ (bottom). The parameters are $A=(2,1.5,1)$,
$\r=(1,0,0)$, $h=80.5$, $l = 12.80$. Incoming intersections of the pink orbit are all on the
left, outgoing on the right.}
\end{minipage} \\[1ex]
\end{figure}

The graphical implementation of this idea is automatically achieved if we first transform to new
polar
coordinates $(\vartheta',\psi')$ where $\ga=\hat\r$ is taken as polar axis instead
of $\ga = (0,0,1)$, and then interpret $\vartheta' = 0$ and $\vartheta'=\pi$ not as points on
a sphere but as circles on a cylinder. An example is shown in Fig.~\ref{Fig:Projections12}
where the two representations
are compared. The top row exhibits S$^\mp$ as in Fig.~\ref{Fig:projpm}, the bottom row T$^\mp$.
The north
pole $(\vartheta,\psi) = (\pi/2,\pi/2)$ is transformed into the circle $\vartheta'=0$ (upper
boundary), and the south pole into the circle $\vartheta'=\pi$ (lower boundary). Identification
of these circles between T$^-$ and T$^+$ produces the full torus T$^2_2$ on which the
non-accessible white regions are now the two holes in the neighborhoods
of the points $(\vartheta',\psi') = (\pi/2,0)$ and $(\pi/2,\pi)$. Since corresponding boundaries
$\partial \Uhl$ on T$^-$ and T$^+$ must also be identified, each pair of holes on the two sides
generates a handle. Hence, the surface of section $\Phl$ for
this example has the topology of a torus with two handles, i.\,e., it is a manifold M$^2_3$.
The lines of tangency, in this particular example,
seem to have become lines of constant $\vartheta'$.

One trajectory has been added, in pink color.
Its incoming part $\dot S < 0$ lies entirely on T$^-$,
the outgoing part $\dot S > 0$ on T$^+$. Note that the incoming part of the
trajectory comes close to the pole $\ga = \hat\r$ whereas the outgoing part has points close to
the south pole $\ga = -\hat\r$. The comparison of upper and lower row illustrates that the
torus T$^2_2$
is the most natural basis for a projection of~$\Phl$.

Simpler manifolds occur, e.\,g., when $\Uhl$ is a topological disk on S$^2(\ga)$. If this disk
does not include any of the poles $\pm\hat\r$, its representation on the PP-torus is two disks, one
on each half. The surface of section is then the union of two disks which are glued together at
their boundary; in other words, $\Phl$ is a topological sphere S$^2$. If the disk contains just
one of the poles, its representation on the torus is a cylinder, and $\Phl$ is the torus T$^2$
obtained by identifying its upper and lower rim. If both poles are contained in the disk, the same
kind of arguments show that $\Phl$ is a 2-manifold of genus~2.

\begin{table}[h]
\begin{tabular}{|c||c|c|c|c|}
\cline{2-5}
\multicolumn{1}{c||}{} & \multicolumn{4}{c|}{Topology of a connected component of
$\Ehl$ \rule[-2mm]{0mm}{7mm}}\\
\cline{2-5}
\multicolumn{1}{c||}{} & $\R$P$^3$ & S$^3$ & S$^1\times$S$^2$ &
(S$^1\times$S$^2$)$\#$(S$^1\times$S$^2$)
 \rule[-2mm]{0mm}{7mm} \\
\cline{2-5}
\multicolumn{1}{c||}{} & \multicolumn{4}{c|}{Structure of $\Uhl$} \rule[-2mm]{0mm}{7mm}\\
\hline $\downarrow$ number of poles in $\Uhl$ & S$^2$ & D$^2$ & S$^2\backslash$2D$^2$ &
                                       S$^2\backslash$3D$^2$  \rule[-2mm]{0mm}{7mm}\\
 \hline
 0 & - & S$^2$ & T$^2$ & M$_2^2$ \rule[-1mm]{0mm}{6mm}\\
 1 & - & T$^2$ & M$_2^2$ & M$_3^2$ \rule[-1mm]{0mm}{6mm}\\
 2 &  T$^2$ & M$_2^2$ & M$_3^2$ & M$_4^2$ \rule[-1mm]{0mm}{6mm}\\
\hline
\end{tabular}
\caption{\label{PSS2:tab1}Possible topologies of the Poincar\'e
surface of section~$\Phl$.}
\end{table}

It is known from the work of Tatarinov~\cite{Tatarinov74} and Bolsinov et al.~\cite{BDW96} that
connected components of $\Uhl \subseteq \text{S}^2(\ga)$ come only in four topological kinds:
the entire sphere S$^2$, a disk D$^2 = \text{S}^2\backslash\text{D}^2$, an annulus
S$^2\backslash\text{2D}^2$, and
a sphere with three holes S$^2\backslash\text{3D}^2$. The corresponding topological types of
$\Phl$ are listed in Tab.~\ref{PSS2:tab1}, for all possible locations of the poles relative
to~$\Uhl$. They are manifolds of genus~$g$ with $g$ ranging from 0 to~4. If we distinguish pairs
of $\Ehl$ and $\Phl$ when either of the partners is different, then the table shows there are 10
different possibilities.

To wrap it up, we propose to start with the PP-torus T$^2_2(\vartheta',\psi')$
as the appropriate manifold onto
which the Poincar\'e surface of section~$\Phl$ is to be projected 1:1. It is
constructed from two
copies of the Poisson sphere where $+\hat\r$ serves as reference point for the polar coordinates
$(\vartheta',\psi')$. The spheres are punctuated at the two poles $+\hat\r$ and $-\hat\r$
which are replaced by circles $(\vartheta',\psi') = (0,\psi')$ and $(\pi,\psi')$, respectively.
Identifying corresponding circles on these two cylinders produces the torus. It will be seen in the
next section that this identification involves a relative shift of the two angles~$\psi'$ by~$\pi$.
The PP-torus carries
two copies of the accessible region~$\Uhl$. Boundaries $\partial\Uhl$ of these two copies are to be
identified. The resulting manifold is a 1:1 image of the complete Poincar\'e surface of
section~$\Phl$.

\section{Examples}
 \label{Sec:PhlStr}

\begin{figure}[H]
\begin{minipage}[t]{\textwidth}

\begin{multicols}{2}

\begin{center}
\includegraphics[width=0.45\textwidth]{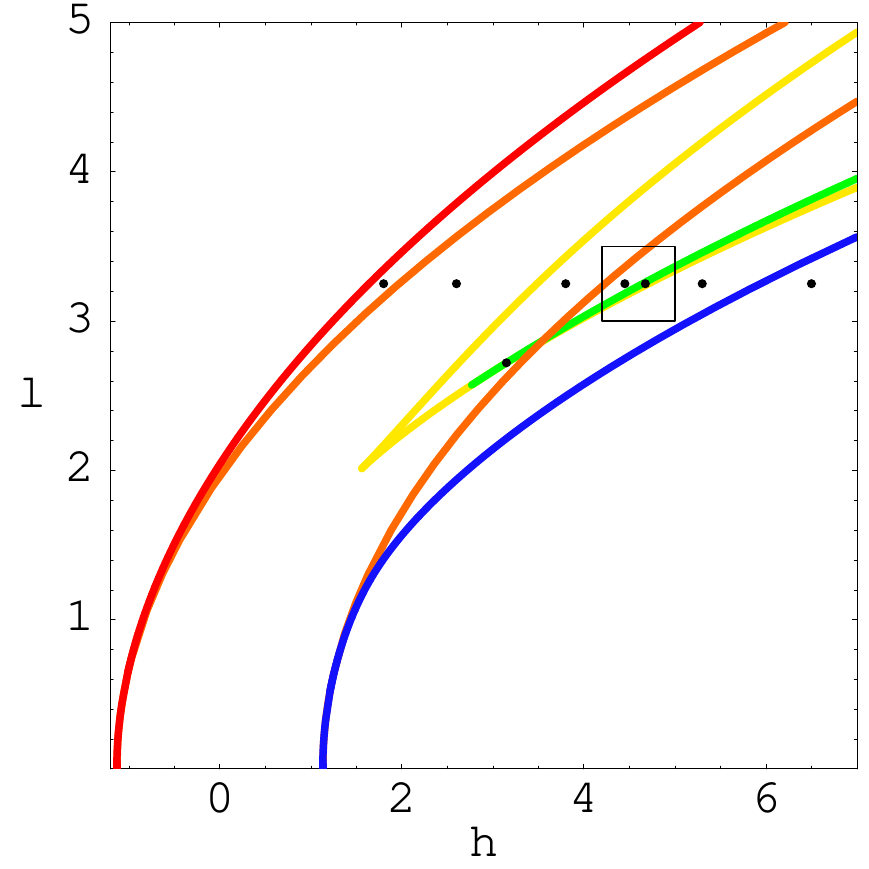}
\end{center}

\newpage

\begin{center}
\includegraphics[width=0.45\textwidth]{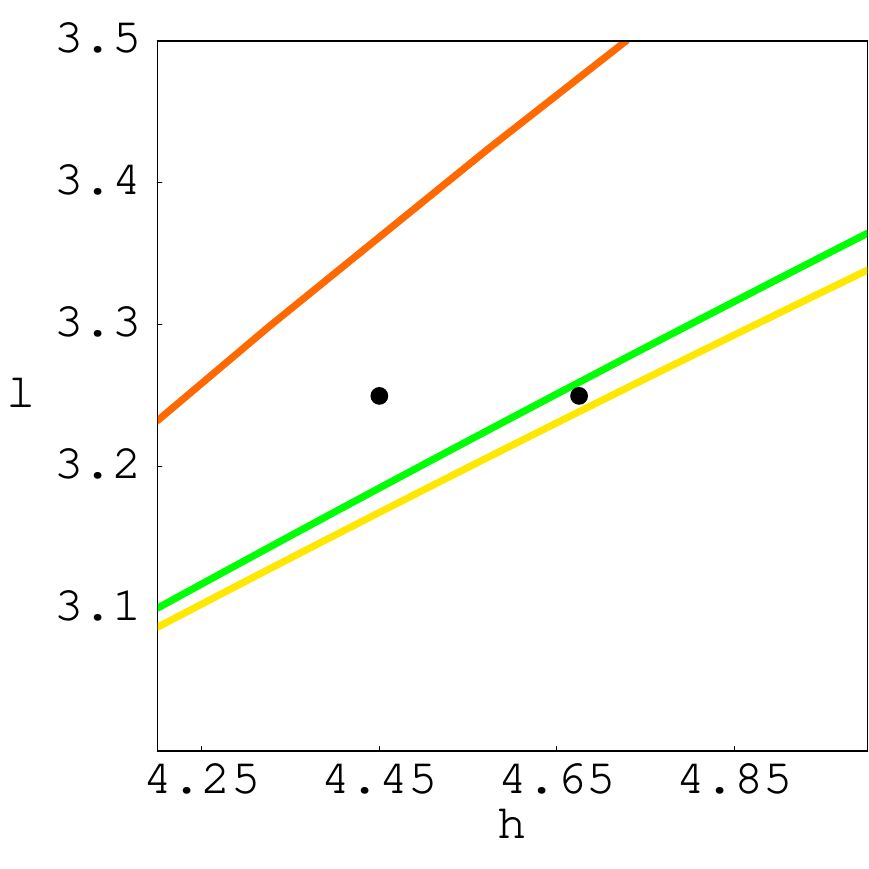}
\end{center}

\end{multicols}

\begin{multicols}{2}

\begin{center}
\includegraphics[width=0.5\textwidth]{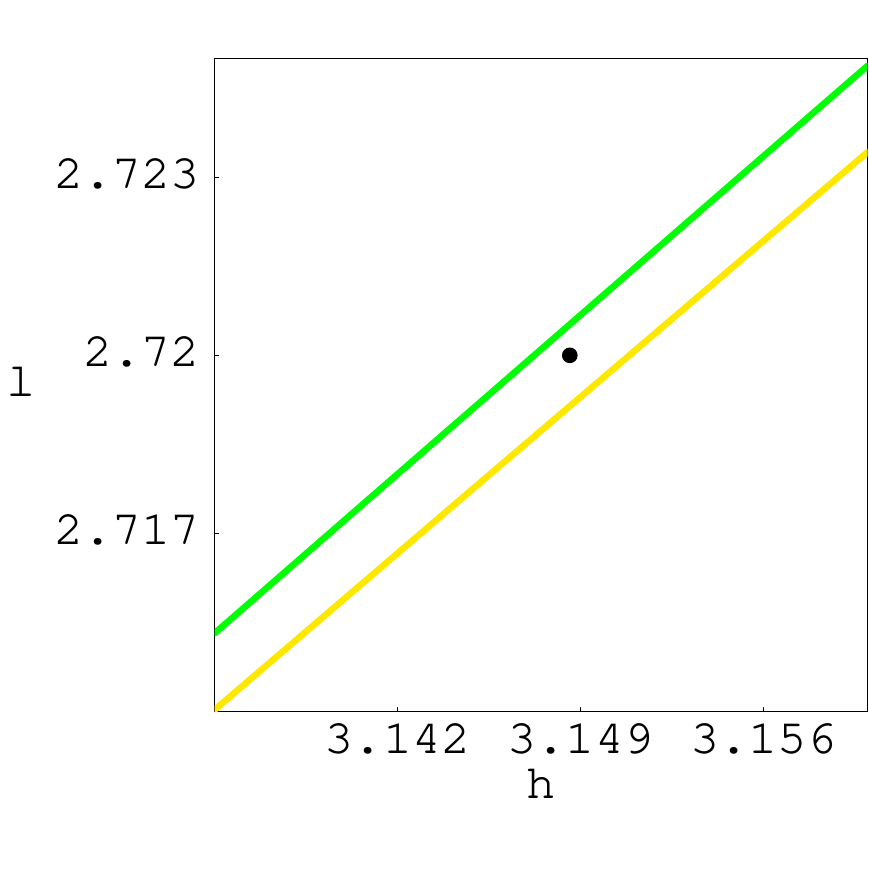}
\end{center}

\newpage

\begin{center}
\end{center}

\end{multicols}

\caption{\label{Fig:BifGen}\small Bifurcation diagram for the Poincar\'e surface of
section~$\Phl$, for $A=(2,1.1,1)$ and $\r = (0.94868,0,0.61623)$. The red, blue, green, and yellow
lines correspond to critical values of~\eqref{EMM:equ1}, the two orange lines come
from~\eqref{Eq:BifPhlr}.
Top right: Magnification of  the rectangle marked on the left. Bottom left: Magnification around the point $(h,l)= (3.1486,2.72)$.
}

\end{minipage}
\end{figure}

In this last section we give nontrivial examples of surfaces of section~$\Phl$ where the extra
bifurcation lines~\eqref{eq:collinear}, or
\begin{equation}
 \label{Eq:BifPhlr}
  h = \frac{l^2}{2\langle \hat\r, A\hat\r \rangle} \mp r \, ,
\end{equation}
do not coincide with bifurcations of the energy surface. This cannot happen in the
Katok family of systems considered in~\cite{GR2004} (which includes
all cases of Lagrange and Kovalevskaya) because there the center of mass~$\r$ lies on a
principal axis, and then~\eqref{Eq:BifPhlr} is the bifurcation line corresponding to the
relative equilibrium where the body rotates about that axis. In order to make~\eqref{Eq:BifPhlr}
different from any of the relative equilibrium conditions, we must choose $\r$ off the
principle axes. An arbitrary example is
shown in Fig.~\ref{Fig:BifGen}, with $A=(2,1.1,1)$ and $\r = (0.94868,0,0.61623)$. The two orange
lines separating from the red and blue relative equilibria correspond to $\ga = \hat\r$ (left)
and $\ga = -\hat\r$ (right).
When $h$ is increased at fixed~$l$, $\Uhl$ is at first
a disk which does not contain any of the poles. Upon crossing the left orange line, $+\hat\r$
enters~$\Uhl$, then at the right orange line, $-\hat\r$ comes in as well.

The six lines in Fig.~\ref{Fig:BifGen} show a partition of the $(h,l)$-plane (with $l>0$) into
eight regions (plus the non-physical region to the left of the red line). Seven of them are
represented by the black dots on the line $l = 3.25$. The tiny eighth region lies between the yellow and
green lines to the left of the orange line (black dot at $(h,l)= (3.1486,2.72)$).
An additional
region, not contained in the figure, exists because at large values of $h$ and $l$ the left branches
of the yellow and orange
lines intersect; point $(h,l)=(14.0,7.2)$ is representative for this region
where the left orange line lies
to the right of the yellow line.

These nine regions correspond to nine different topologies of~$\Phl$, including six of the types listed
in Tab.~\ref{PSS2:tab1}, and three examples of disjoint surfaces of section. We discuss them one by
one. The following Poincar\'e sections contain typical orbits, with different colors for
different orbits. When certain orbits are the same type (elliptic or hyperbolic) in different regions
they are given the same color.

\medskip

\begin{figure}
\begin{center}
\includegraphics[width=0.3\textwidth]{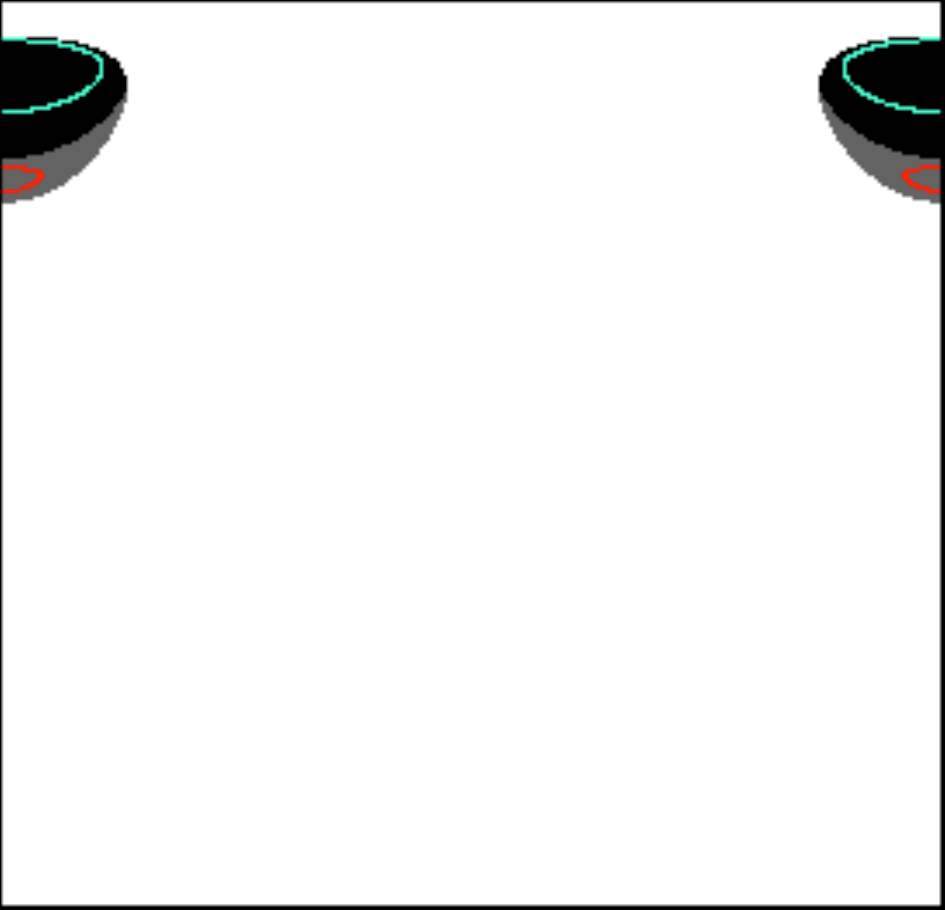}
\hspace{1cm}
\includegraphics[width=0.3\textwidth]{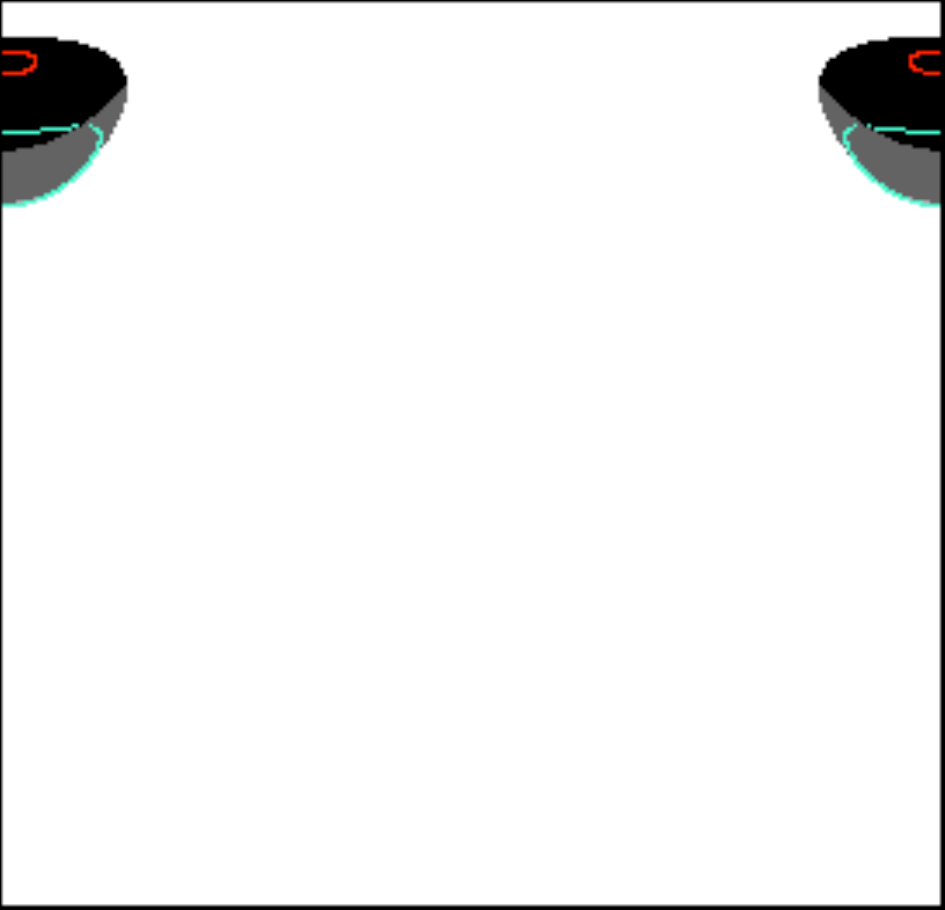}
\end{center}
\begin{center}
\includegraphics[width=0.3\textwidth]{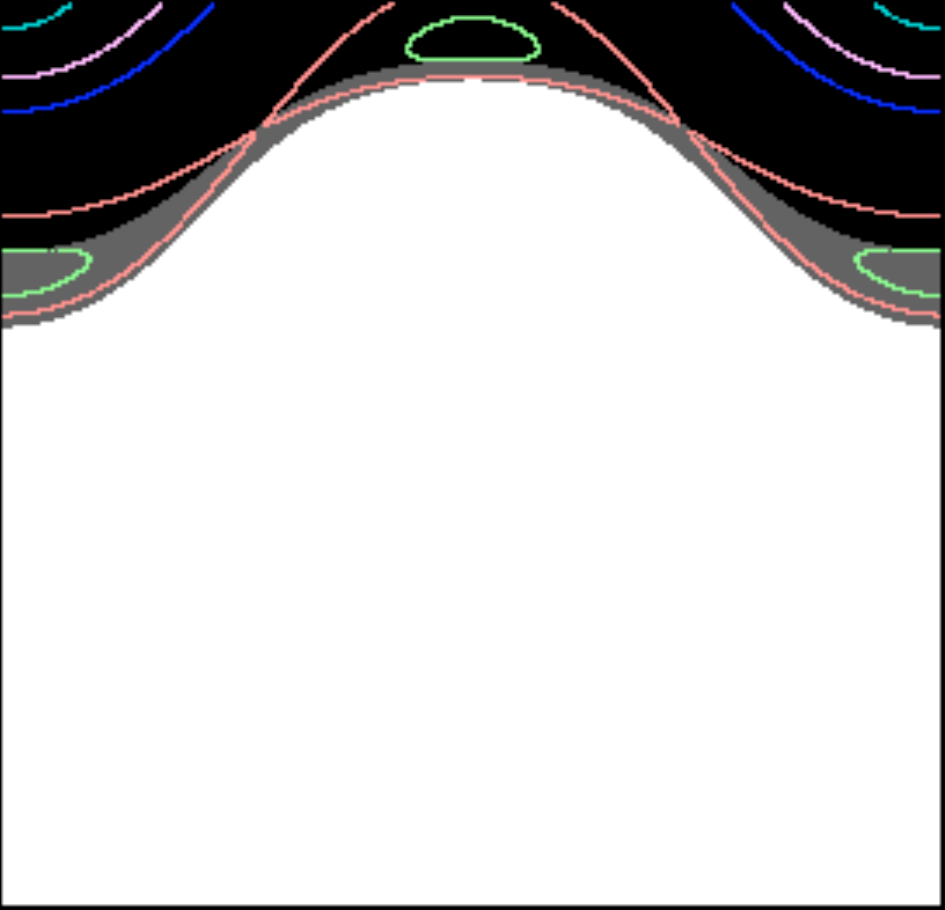}
\hspace{1cm}
\includegraphics[width=0.3\textwidth]{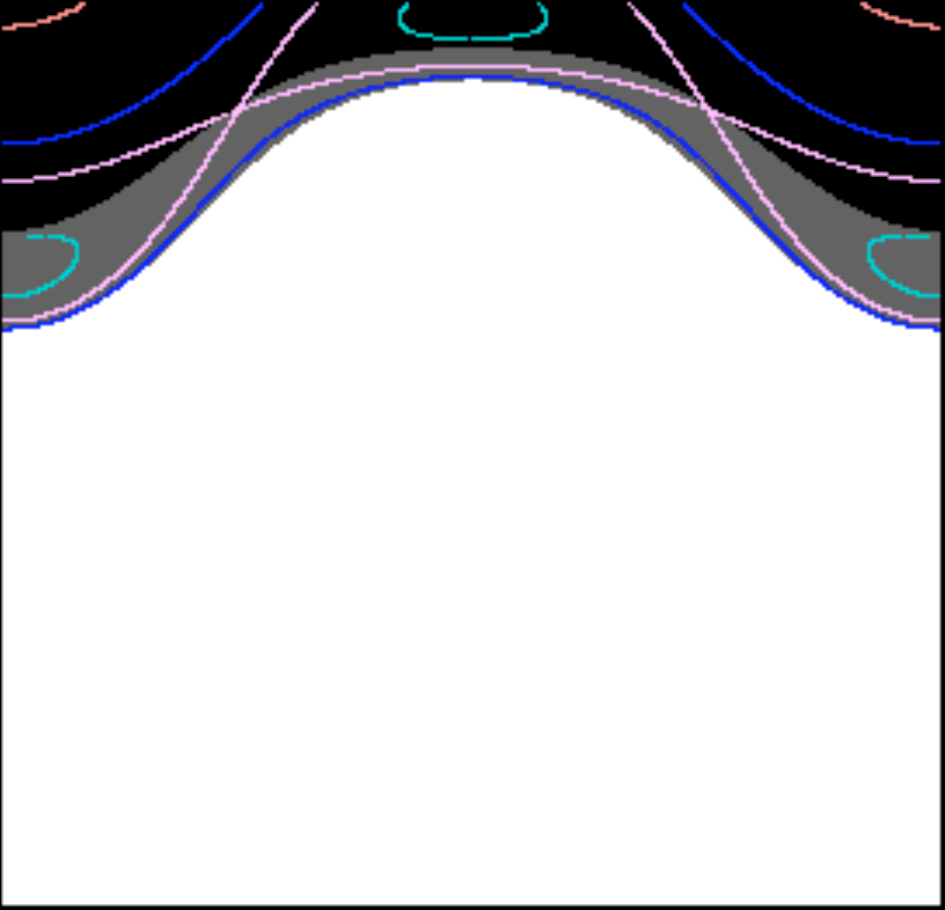}
\end{center}
\begin{center}
\includegraphics[width=0.3\textwidth]{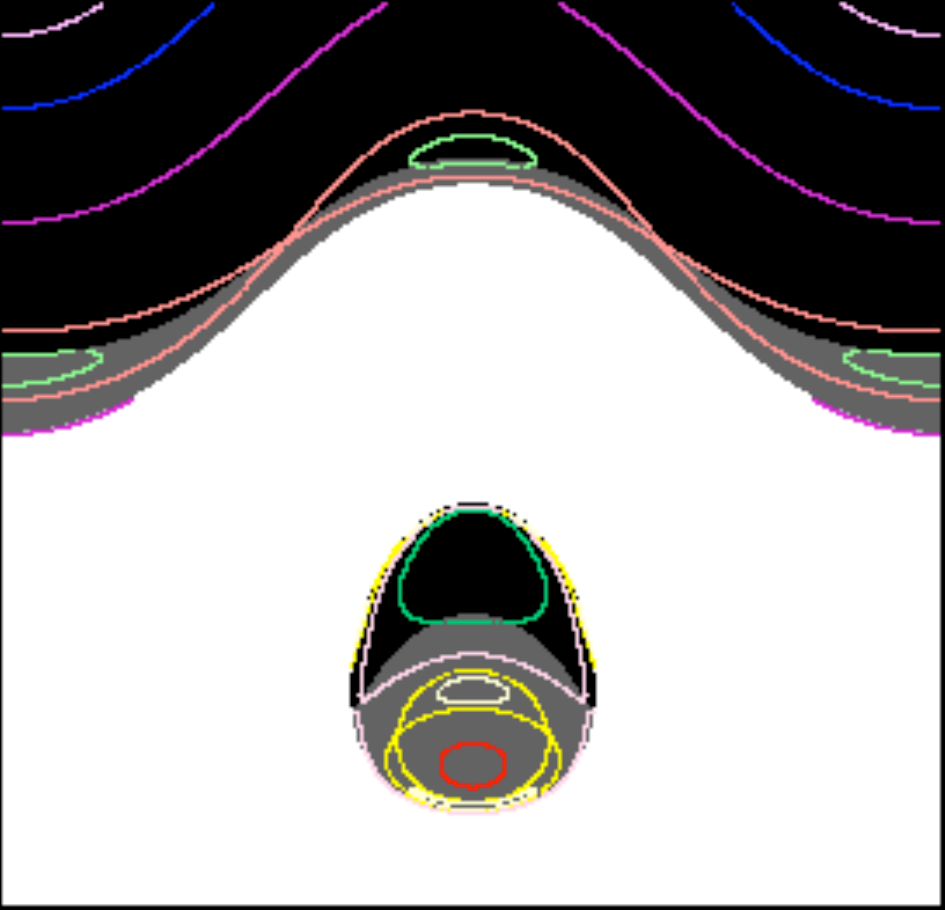}
\hspace{1cm}
\includegraphics[width=0.3\textwidth]{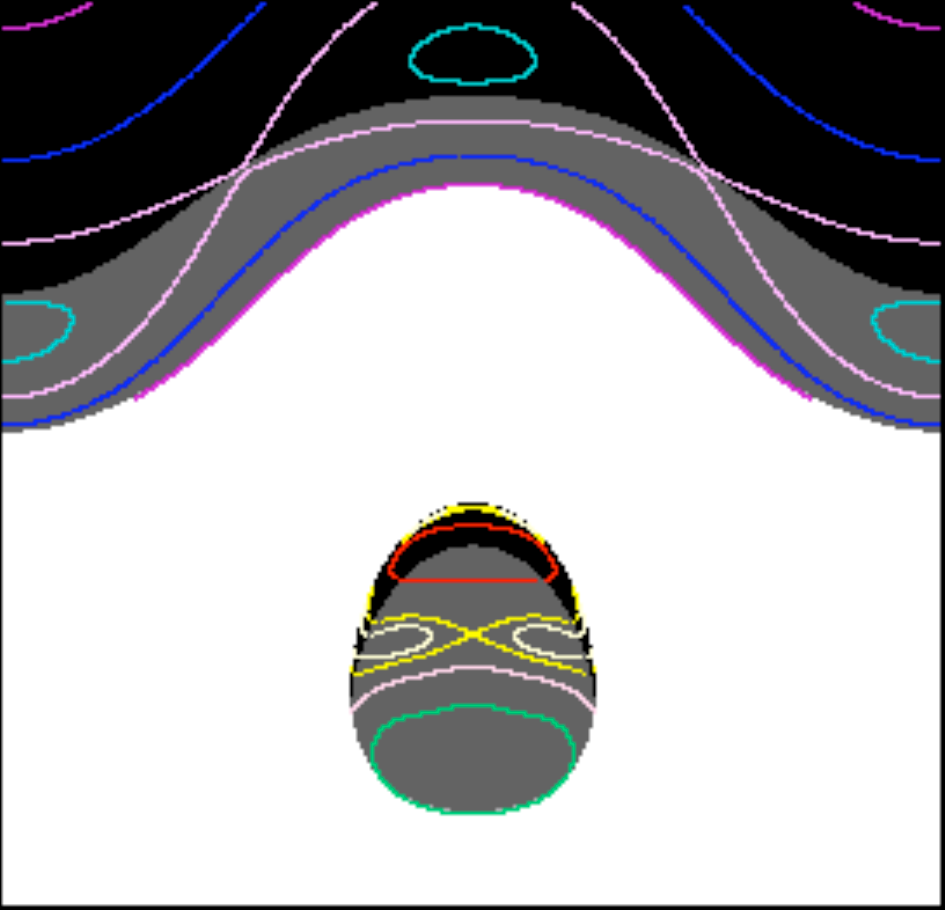}
\end{center}
\begin{center}
\includegraphics[width=0.3\textwidth]{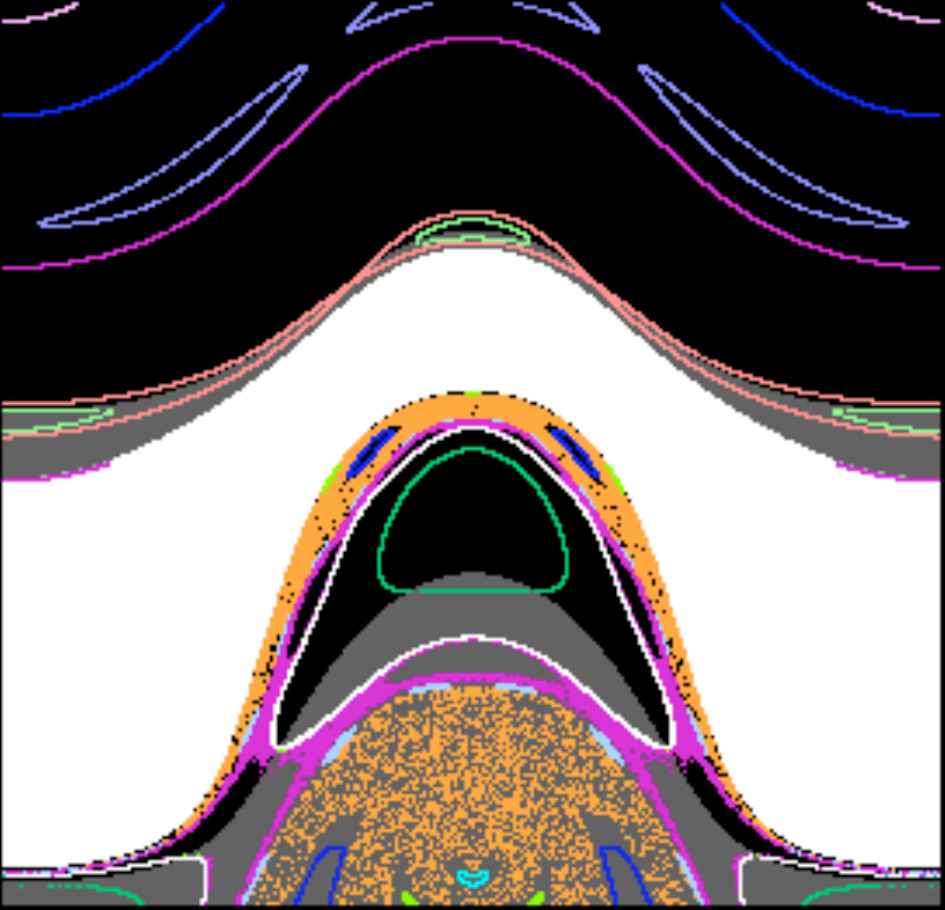}
\hspace{1cm}
\includegraphics[width=0.3\textwidth]{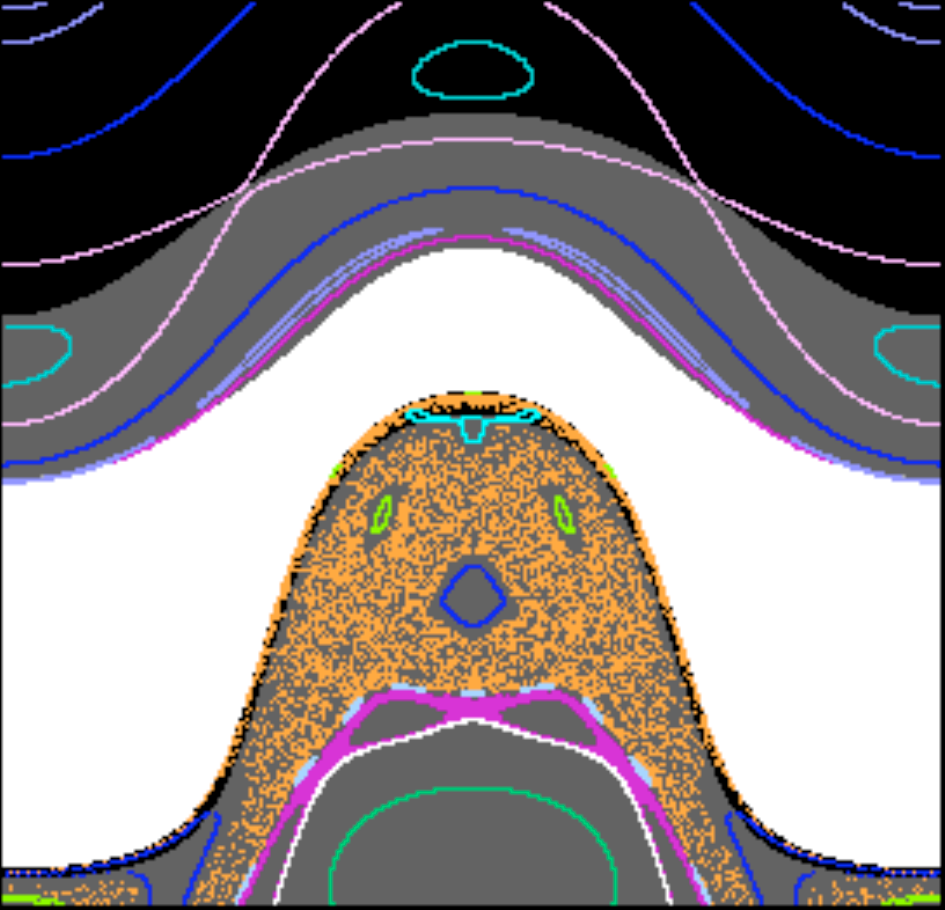}
\end{center}
\caption{\label{Fig:PhlStr1}\small Projections of $\Phl$ onto T$^-$ (left) and T$^+$ (right).
Parameters are those of
Fig.~\ref{Fig:BifGen}. The angular momentum is $l=3.25$ for all cases, the energy
from top to bottom is $1.8$, $2.6$, $3.8$ and $4.45$.}
\end{figure}

The first example $(h,l) = (1.8,3.25)$ represents the leftmost region between the red and orange
bifurcation lines. It is shown in the top row of Fig.~\ref{Fig:PhlStr1}. The accessible region
$\Uhl$ on T$^{\mp}$ is a rather small topological disk which does not contain the center of
mass~$\hat{\r}$. Gluing the two copies together along the boundary of $\Uhl$ we obtain for $\Phl$ a
topological sphere~S$^2$.

The second row of Fig.~\ref{Fig:PhlStr1} shows the case $(h,l)=(2.6,3.25)$. The north pole (upper
rim) has entered $\Uhl$. The orbit structure reveals that when we identify the two circles
corresponding to $\hat{\r}$ on T$^{\mp}$, the angle $\psi'$ must be shifted by~$\pi$. On the
Poisson sphere, $\Uhl$ is again a disk D$^2$, but on the torus $\text{T}^+ \cup \text{T}^-$ its two
copies together form an annulus.
Gluing together the two boundaries $\partial \Uhl$ on T$^2_2$ we find
that $\Phl$ is a torus T$^2$.

Increasing the energy beyond the yellow line in Fig.~\ref{Fig:BifGen},
the topology of $\Ehl$ changes from S$^3$ to 2S$^3$,
a new disjoint disk appears as part of $\Uhl$. This disk contains no pole, hence we
deduce that $\Phl$ is the union of a torus T$^2$ and a sphere S$^2$, see
the third row of Fig.~\ref{Fig:PhlStr1} with $(h,l)=(3.8,3.25)$. It may be surprising that
the trajectories shown indicate only regular behavior with no sign of chaos even though
the equations of motion are certainly non-integrable.

\clearpage

\clearpage

The next point $(h,l)=(4.45,3.25)$, see the right part of Fig.~\ref{Fig:BifGen} and the last row
of Fig.~\ref{Fig:PhlStr1}, represents a region where both disjoint disks of $\Uhl$ contain a pole.
The surface of section is therefore 2T$^2$. It appears that the accessibility of the south pole
(corresponding to the center of mass in upright position) suddenly introduces manifest chaos. While
the component around $\hat{\r}$ still looks very regular, the one containing $-\hat{\r}$ is
dominated by chaotic motion. The orange and purple orbits belong to the same chaotic region but
for a long time remain separate, indicating the presence of cantori related to the chain of islands
between them.

\begin{figure}
\begin{center}
\includegraphics[width=0.3\textwidth]{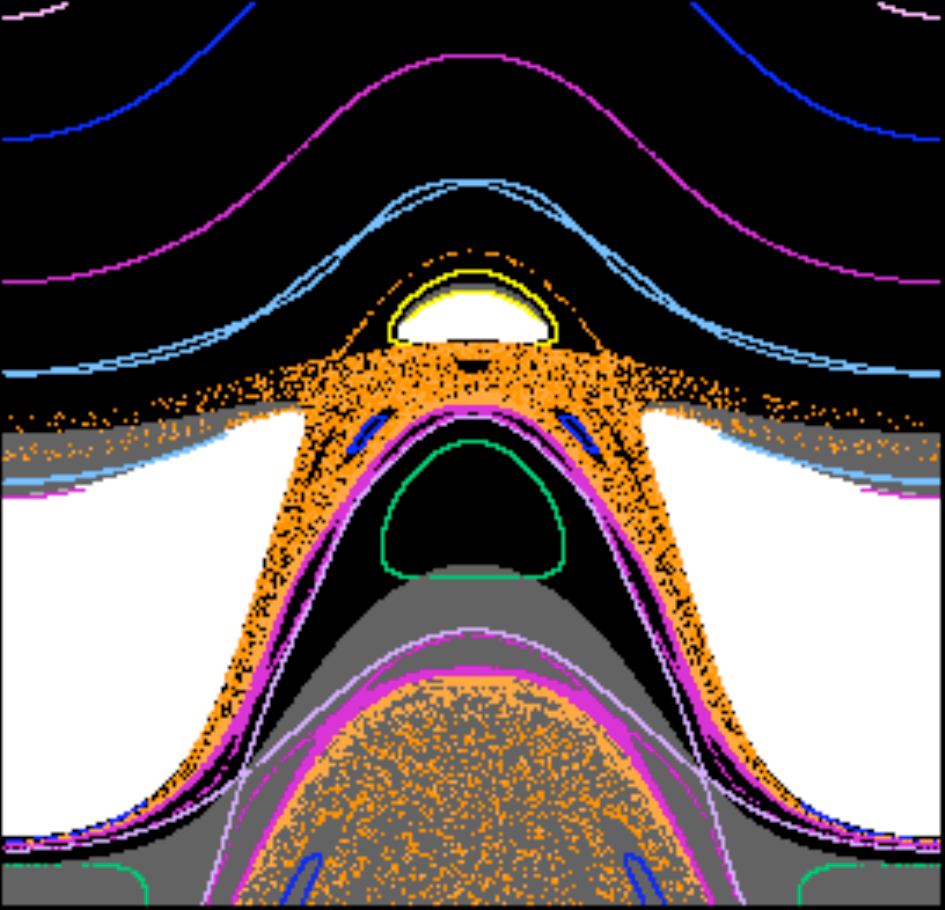}
\hspace{1cm}
\includegraphics[width=0.3\textwidth]{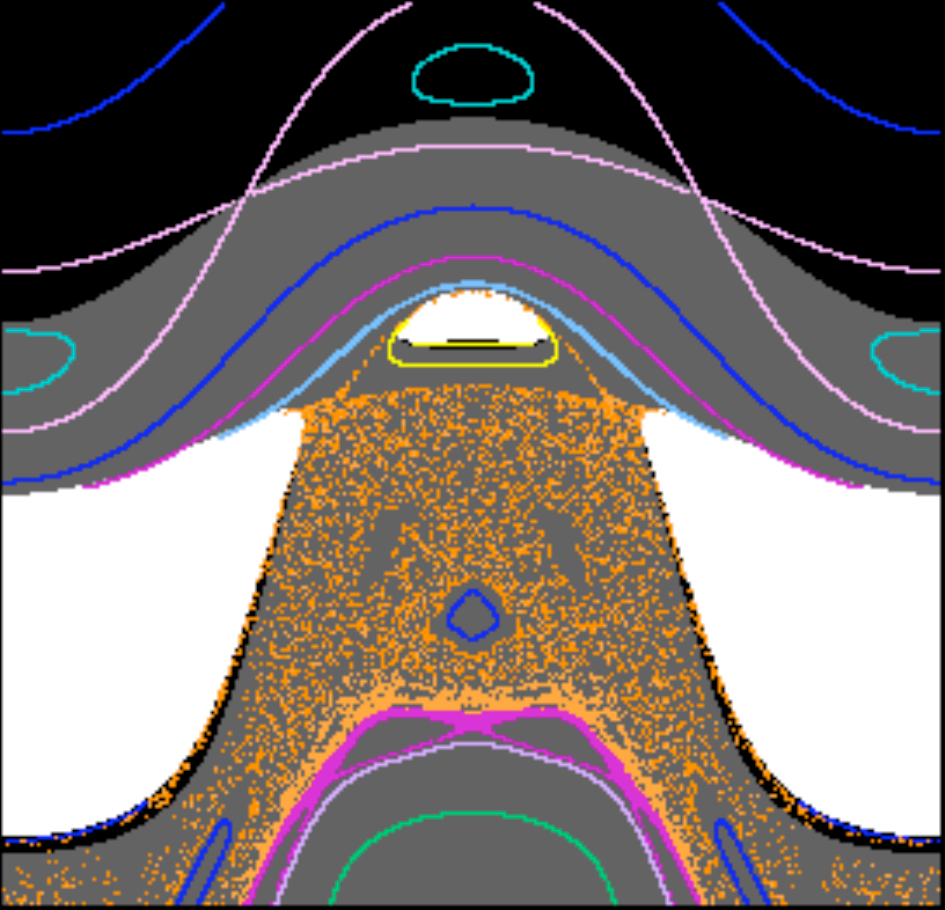}
\end{center}
\begin{center}
\includegraphics[width=0.3\textwidth]{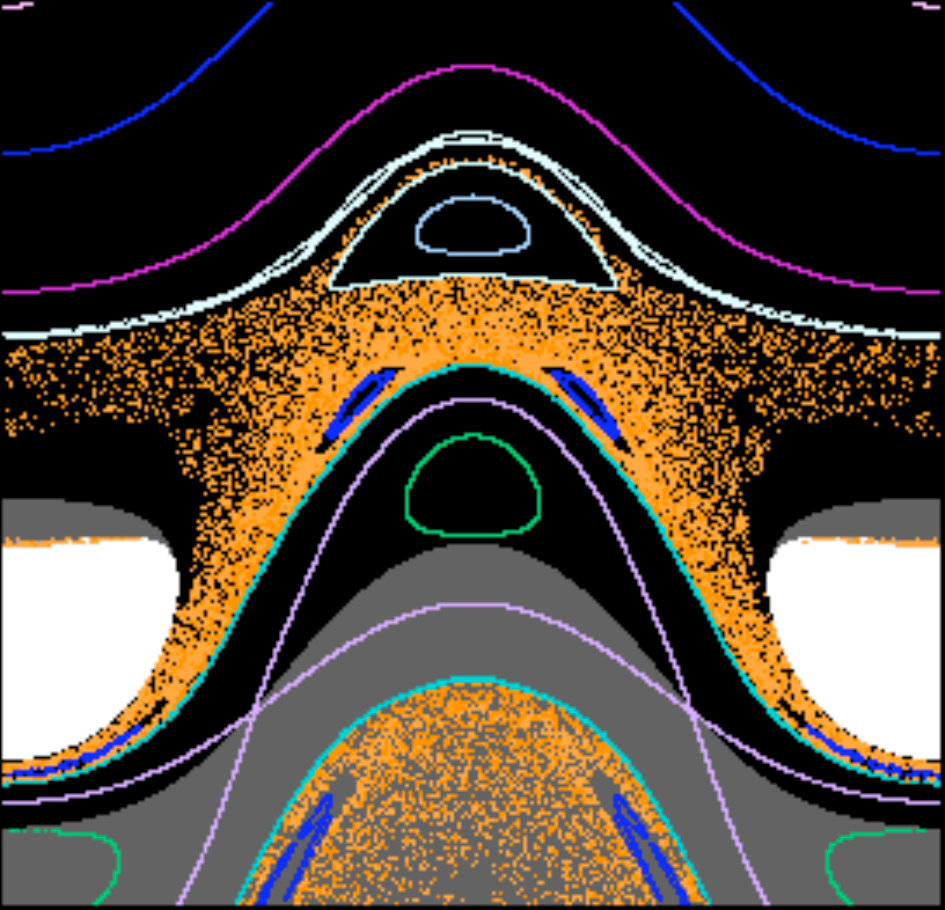}
\hspace{1cm}
\includegraphics[width=0.3\textwidth]{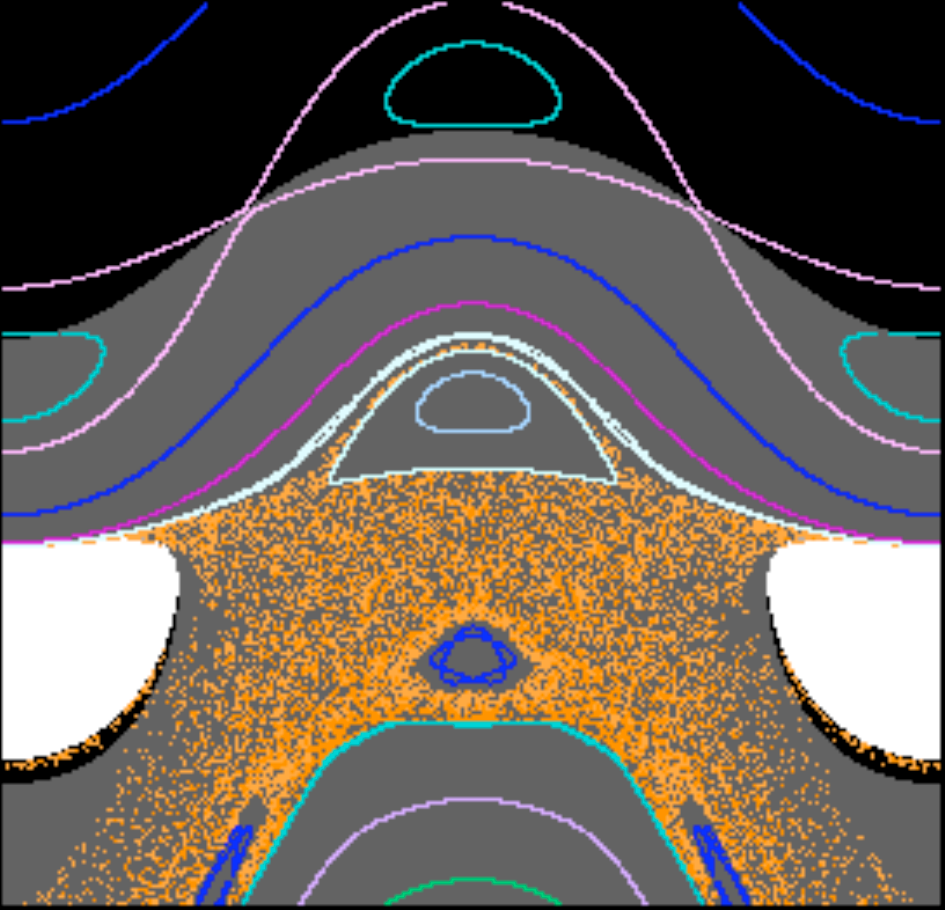}
\end{center}
\begin{center}
\includegraphics[width=0.3\textwidth]{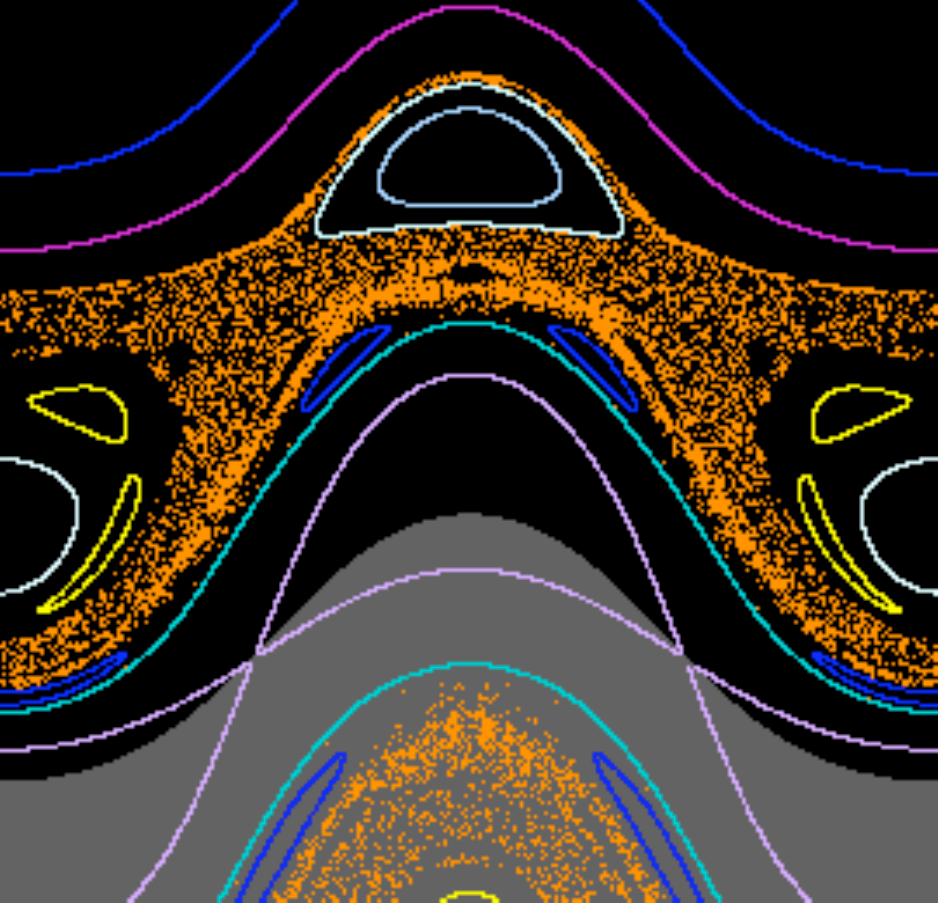}
\hspace{1cm}
\includegraphics[width=0.3\textwidth]{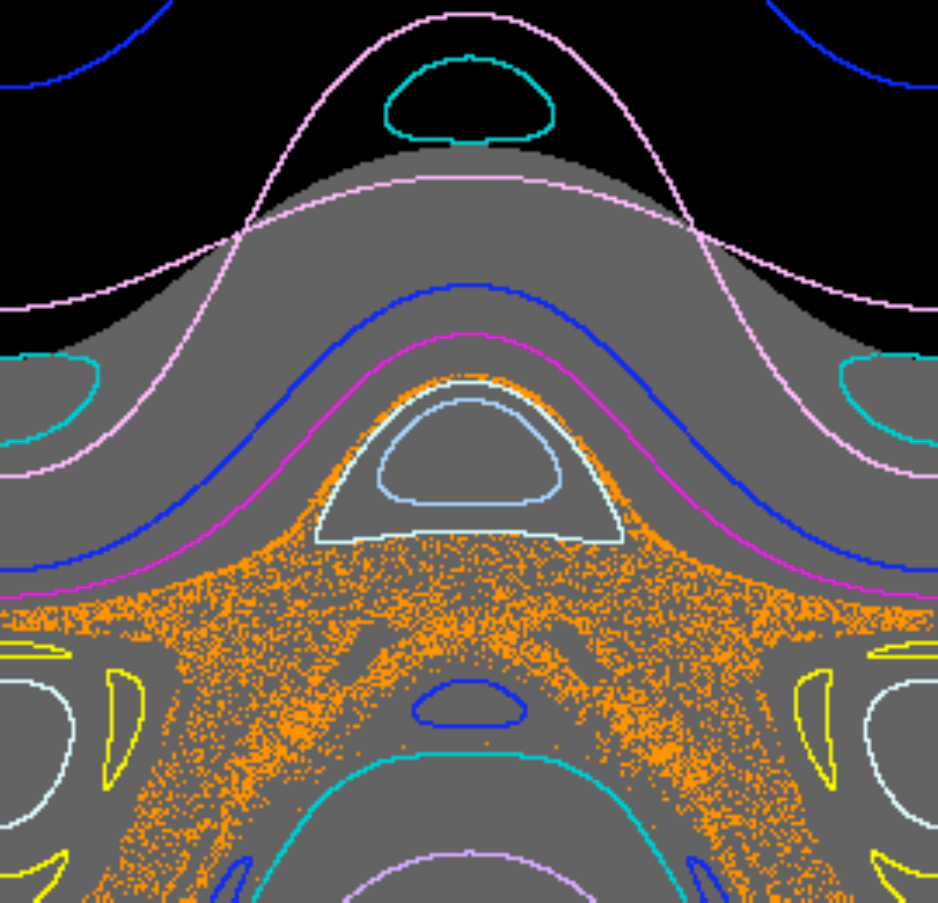}
\end{center}
\caption{\label{Fig:PhlStr2}\small Projections like in Fig.~\ref{Fig:PhlStr1}.
Again $l=3.25$ in all three cases; the energy from top to bottom is $4.675$, $5.3$ and $6.5$.}
\end{figure}

The next three cases with $l=3.25$ are shown in Fig.~\ref{Fig:PhlStr2}. The first row with
$h=4.675$ represents the narrow $(h,l)$-region between the green and yellow bifurcation lines,
see the right part of Fig.~\ref{Fig:BifGen}. $\Uhl$ is the Poisson sphere minus
two disks, i.\,e., a topological annulus; as this annulus contains both poles, the surface of section $\Phl$
is a manifold M$_{3}^2$ of genus~3. Compared to the previous case $h = 4.45$,
the two disks about the poles have merged at two points and now allow the chaos to sneak into the
upper region. Nevertheless, the neighborhood of $\hat{\r}$ remains mostly regular.

\weglassen{
Although there is a gap between the upper and lower region they are still separated by invariant
curves. (Das verstehe ich nicht: was fuer ein "gap"?)
So the chaos cannot spread into the region close to the north pole. Note that the chaos
band has now split into two disconnected pieces. Here, too, we note a phase shift in $\psi$ of
$\pi$. The overall structure of the invariant curves remains mostly the same, however.
}

One of the two inaccessible disks disappears as $h$ is increased beyond the yellow line in
Fig.~\ref{Fig:BifGen}, see the second row of Fig.~\ref{Fig:PhlStr2} where $h=5.3$. $\Uhl$ is now a
disk which contains both poles, hence $\Phl$ is a manifold M$_{2}^{2}$ of genus~2. It appears
that the phase space is fairly distinctly divided into one chaotic and two regular parts. The
``outgoing'' intersections of the chaotic orbits take place in a region connected to the south
pole~$\hat{\r}$, the ``incoming'' lie between the poles. A bigger regular region surrounds the
north pole, a smaller one connects to the south pole at about one half of the $\psi'$-circle.
So chaos and order meet at the south pole, depending on the angle~$\psi'$.

Finally, when $h$ crosses the blue bifurcation line, the whole Poisson sphere becomes accessible.
This is shown, with $h=6.5$, in the last row of Fig.~\ref{Fig:PhlStr2}. The surface of section
is now isomorphic to the entire PP-torus T$^2_2$. As to
the distribution of regular and chaotic orbits on this torus, we observe again two regular parts
separated by two chaotic regions.

\clearpage

\begin{figure}
\begin{center}
\includegraphics[width=0.3\textwidth]{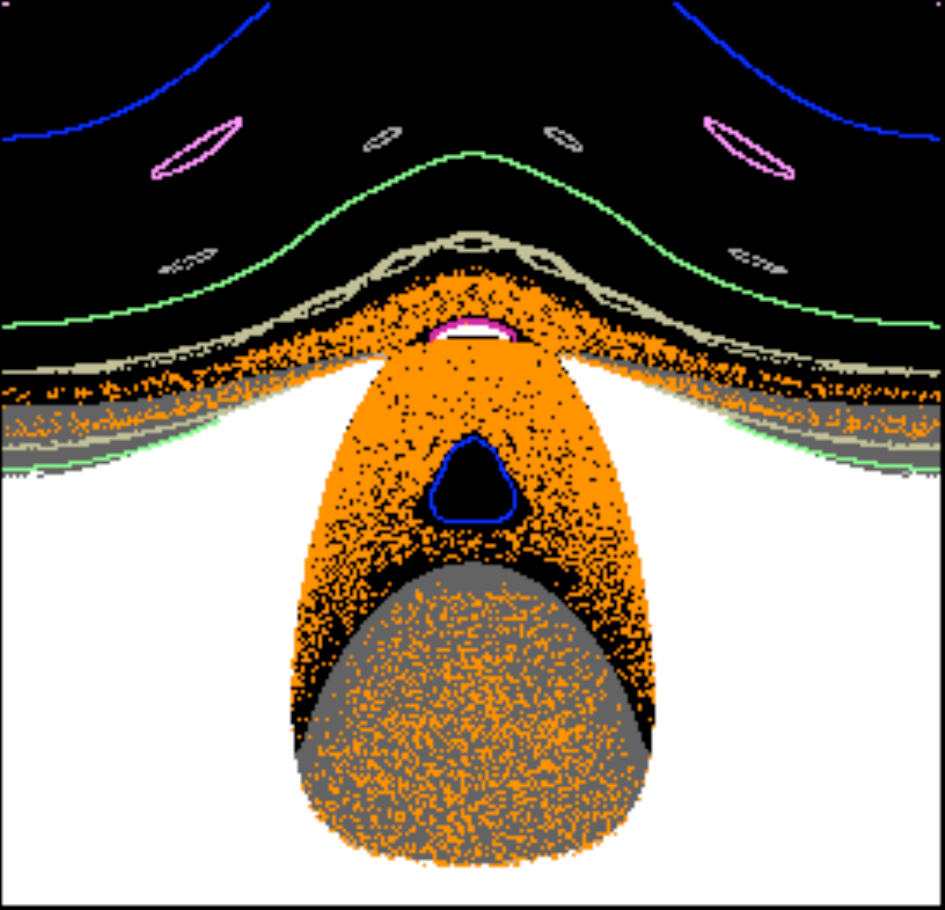}
\hspace{1cm}
\includegraphics[width=0.3\textwidth]{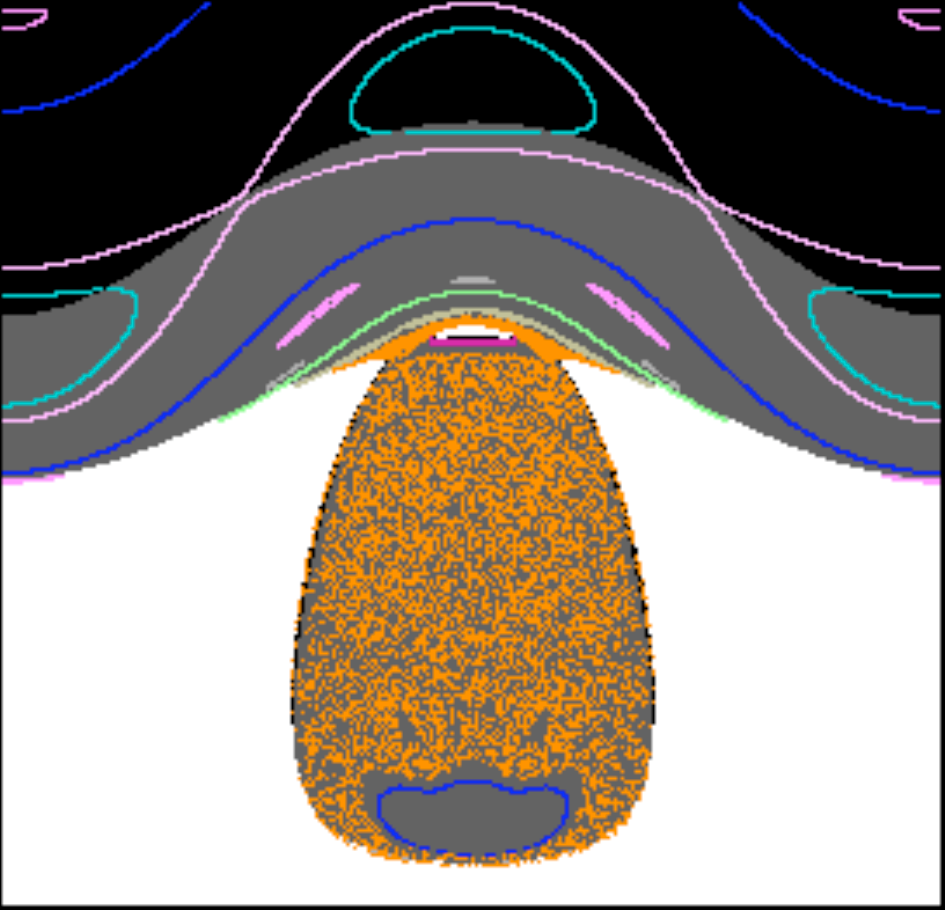}
\end{center}
\begin{center}
\includegraphics[width=0.3\textwidth]{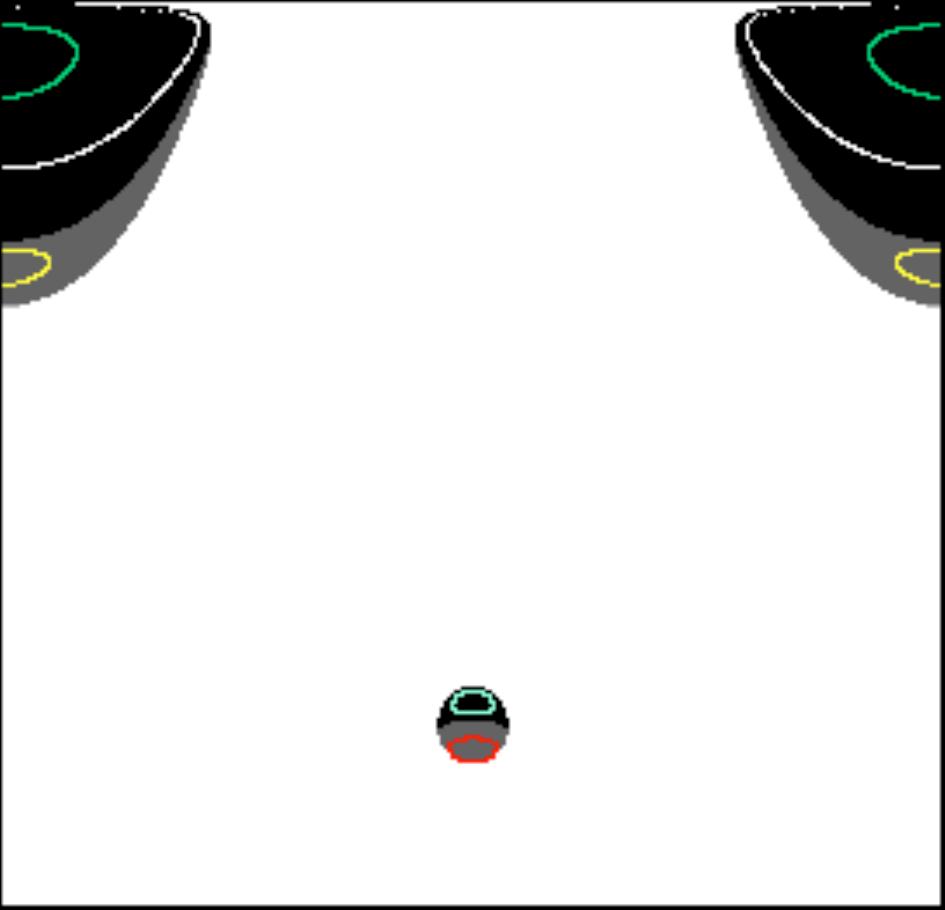}
\hspace{1cm}
\includegraphics[width=0.3\textwidth]{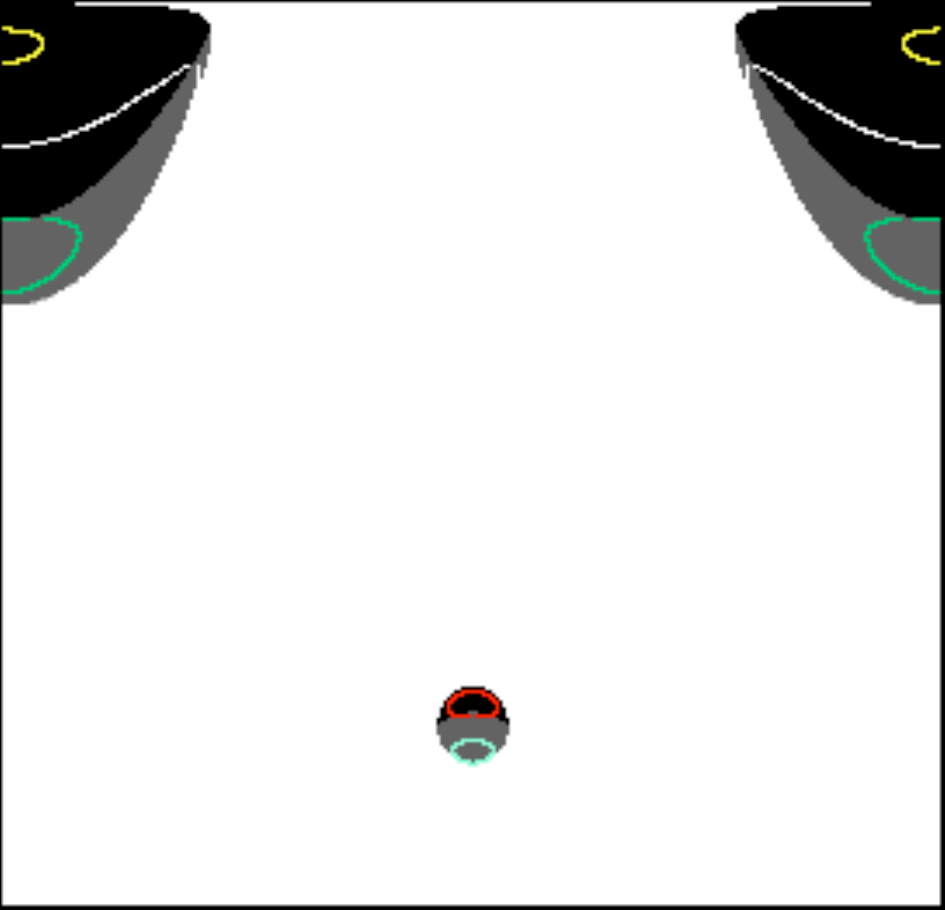}
\end{center}
\caption{\label{Fig:PhlStr3}\small Projections of $\Phl$ like in Fig.~\ref{Fig:PhlStr1}.
Top row $(h,l)=(3.1486,2.72)$, bottom row $(14.0,7.2)$ (bottom).}
\end{figure}

Two more types of surfaces of section are shown in Fig.~\ref{Fig:PhlStr3}. They do not exist
for $l=3.25$. The upper row shows the example $(h,l)=(3.1486,2.72)$; the corresponding dot in
Fig.~\ref{Fig:BifGen} lies between the green and yellow lines, but to the left of the orange
line. Compared to the first row of Fig.~\ref{Fig:PhlStr2}, the lower part does not contain
the south pole, yet there remains a small hole (barely visible in the figure) inside the accessible
region, near the center. $\Uhl$ is a topological annulus containing the north pole, hence
$\Phl$ is of type M$_{2}^{2}$. In the scenario where this case emerges from
the third row of Fig.~\ref{Fig:PhlStr1}, an outbreak of chaos takes place in the lower part.

The last case is $(h,l)=(14.0,7.2)$; the accessible region consists of two disks neither of which
contains a pole. Hence $\Phl$ is the union of two topological spheres.

\clearpage

With this series of nine types of $\Uhl$ we have illustrated 6 out of the 10 possible topologies
listed in Tab.~\ref{PSS2:tab1}. But notice that our choice $A=(2,1.1,1)$ represents only one of the
seven different classes of moments of inertia that Katok distinguishes with $\hat{\r}=(1,0,0)$
(the class K2 in the notation of~\cite{GR2004}). With other choices it is possible to also find
examples for the other types of $\Uhl$.

\section{Summary and outlook}

We have presented a new tool for the study of rigid body dynamics: a complete Poincar\'e surface of
section~$\Phl$ and its 1:1 representation in projection to a suitably defined torus, the
PP-torus~T$^2_2$. The surface $\Phl \subset \Ehl$ is defined by local extrema of the vertical
component of the center of mass~$\r$, $\d \langle \ga,\r\rangle /\d t = 0$, and the PP-torus is
obtained from two copies S$^+(\ga)$ and S$^-(\ga)$ of the Poisson sphere by the following procedure:
punctuate S$^\pm(\ga)$ at the points $\ga = \pm \r/r$, replace these points by ``polar circles''
to obtain two cylinders T$^\pm(\ga)$,
and then identify corresponding polar circles on T$^+(\ga)$ and T$^-(\ga)$.
The torus T$^2_2$ so constructed contains
two copies of the accessible $\ga$-region, and $\Phl$ projects 1:1 to the union of these two copies.
Identifying the boundaries $\partial\Uhl$ on T$^+(\ga)$ and T$^-(\ga)$ produces a manifold which is
homeomorphic to $\Phl$ and at the same time readily accessible to intuition.

We have shown examples for nine topologically different surfaces~$\Phl$ and propose to use the
tool for further exploration of the dynamics of rigid bodies in their 4-dimensional parameter space.
In a forthcoming publication we will apply it to the family of systems $A = (2,2\eta,1)$,
$\r = (1,0,0)$, which contains the integrable system of Lagrange ($\eta=1$) and Kovalevskaya
($\eta=2$). Pictures like those of Figs.~\ref{Fig:PhlStr1}-\ref{Fig:PhlStr3} reveal the fate of
particular features of the phase space structure under parameter variation, such as location and
stability of isolated periodic orbits, their bifurcation schemes, as well as the extent and
entanglement of regular and chaotic regions.

\numberwithin{equation}{section}

\appendix
\section{Appendix: Explicit characterization of $\Phl$}

Given an energy $h$, an angular momentum $l$, and a point $\ga$ on the Poisson sphere
(not collinear with $\r$),
we determine the corresponding vectors~$\l$ on the Poincar\'e surface defined by
$S=0$, or $\langle \l, A^{-1}(\r\times\ga)\rangle = 0$.
For abbreviation we introduce the notation
\begin{equation}
 \u :=  A^{-1}(\r\times\ga), \qquad \v := \u \times \ga \, .
\end{equation}
The equations $S = \langle \l,\u \rangle = 0$ and $L = \langle \l,\ga \rangle = l $ define
planes in $\l$-space which intersect in a line with direction~$\v$. It is easy to check that
\begin{equation}
 \label{eq:l0}
 \l_0 = l\,\frac{\v\times\u}{\langle \v,\v \rangle}
\end{equation}
is one point on this line, hence we may parameterize it with $\l = \l_0 + \v t$.
The energy equation~\eqref{EM:equ2} determines the possible values of~$t$. We write it in the form
\begin{equation}
  \langle \l_0 + \v t, A^{-1}(\l_0 + \v t) \rangle = 2(h + \langle \r,\ga \rangle) =: c \, .
\end{equation}
This gives a quadratic equation for $t$,
\begin{equation}
  \langle \v, A^{-1}\v\rangle t^2 + 2\langle \l_0, A^{-1}\v\rangle t + \langle\l_0,
  A^{-1}\l_0\rangle - c = 0 \, .
\end{equation}
Inserting its solutions into $\l = \l_0 + \v t$, we find $\l = \l_1 \pm \l_2$ with
\begin{equation}
 \l_1 = \l_0 - \frac{\langle \l_0,A^{-1}\v \rangle}{\langle \v,A^{-1}\v \rangle}\,\v
\end{equation}
and
\begin{equation}
 \label{eq:l2}
 \l_2 = \frac{\v}{\langle \v,A^{-1}\v \rangle}\,\sqrt{\langle \l_0,A^{-1}\v \rangle^2 -
       \bigl( \langle \l_0,A^{-1}\l_0 \rangle - c \bigr) \langle \v,A^{-1}\v \rangle } .
\end{equation}
To evaluate further, we use an identity which holds for arbitrary symmetric matrices~$A$ and
vectors $\a$, $\b$:
\begin{equation}
 \label{eq:AaAb}
  A\a \times A\b = (\text{det}A)\,A^{-1}(\a\times\b).
\end{equation}
For example, with $\v = \u\times\ga$ we find $A^{-1}\v = (A\u \times A\ga)/\text{det} A$ and
\begin{equation}
  \langle \v,A^{-1}\v \rangle  = \frac{1}{\text{det} A} \langle \u\times\ga , A\u \times
  A\ga \rangle
   = \frac{1}{\text{det} A} \langle \u , A\u \rangle \langle \ga , A\ga \rangle \, ;
\end{equation}
in the last step we used Lagrange's identity and $\langle \ga, A\u \rangle =
\langle \ga, \r\times\ga \rangle = 0$. With similar arguments, and using the explicit
form~\eqref{eq:l0}
for $\l_0$, we find
\begin{equation}
 \l_1 = l \,\frac{A\ga}{\langle \ga, A\ga \rangle} \, .
\end{equation}
To evaluate $\l_2$, we first employ Lagrange's identity to obtain
\begin{equation}
  \langle \l_0,A^{-1}\v \rangle^2 -
       \langle \l_0,A^{-1}\l_0 \rangle \langle \v,A^{-1}\v \rangle =
  \langle \l_0\times \v , A^{-1}\v \times A^{-1}\l_0 \rangle     \, ,
\end{equation}
and then~\eqref{eq:AaAb} with $A^{-1}$ instead of $A$ to transform this into
\begin{equation}
 \frac{-1}{\text{det} A}\,\langle \l_0\times \v , A(\l_0\times \v) \rangle \, .
\end{equation}
An easy calculation using $\langle \v,\u \rangle = 0$ shows that $\l_0\times\v = l\u$. Inserting
this into the radicand of~\eqref{eq:l2} we find that it may be written as
\begin{equation}
  \frac{\langle \u,A\u \rangle}{\text{det} A}(c\langle \ga, A\ga \rangle - l^2) =
  2\langle \v,A^{-1}\v \rangle \bigl(h - U_l(\ga)\bigr)
\end{equation}
with the effective potential of Eq.~\eqref{EM:equ10}. The final result for $\l$ is
\begin{equation}
 \l = l\,\frac{A\ga}{\langle \ga, A\ga \rangle} \pm \v\,\sqrt{\frac{2(h-U_l(\ga))}
 {\langle \v,A^{-1}\v \rangle }}\, .
\end{equation}


\section*{Acknowledgements}

We thank Igor Gashenenko for many discussions.


\bibliographystyle{plain}
\bibliography{SchmidtDR}


\end{document}